\documentclass[%
 reprint,
 amsmath,amssymb,
 aps,
]{revtex4-2}

\usepackage{graphicx}
\usepackage{dcolumn}
\usepackage{bm}

\usepackage[final]{hyperref}
\usepackage{float}
\hypersetup{
	colorlinks=true,       
	linkcolor=blue,        
	citecolor=blue,        
	filecolor=magenta,     
	urlcolor=blue         
}
\usepackage{braket}
\usepackage{cleveref}
\usepackage{bbm}
\usepackage{tikz}
\usepackage{slashed}
\usepackage{verbatim}
\usepackage[normalem]{ulem}

\usepackage{pgfplots}
\pgfplotsset{compat=newest}
\usepgfplotslibrary{groupplots}
\usepackage{url}
\usepackage{tikz}
\usetikzlibrary{decorations.markings}
\usetikzlibrary{shapes.misc}
\tikzset{cross/.style={cross out, draw=black, minimum size=2*(#1-\pgflinewidth), inner sep=0pt, outer sep=0pt},
cross/.default={1pt}}

\begin{document}

\preprint{APS/123-QED}

\title{Proposal to use laser-accelerated electrons to probe the axion-electron coupling}

\author{Georgios Vacalis$^{1}$}
\email{georgios.vacalis@chch.ox.ac.uk}
\author{Atsushi Higuchi$^{2}$}
\author{Robert Bingham$^{3,4}$}
\author{Gianluca Gregori$^{1}$}

\affiliation{ \small\it $^1$Department of Physics, University of Oxford, Parks Road, Oxford OX1 3PU, UK}
\affiliation{ \small\it$^2$ Department of Mathematics, University of York, Heslington, York YO10 5DD, UK}
\affiliation{\small\it $^3$Rutherford Appleton Laboratory, Chilton, Didcot, Oxon OX11 OQX, UK,}%

\affiliation{\small\it $^4$Department of Physics, University of Strathclyde, Glasgow G4 0NG, UK}%
\date{\today}

\begin{abstract}
The axion is a hypothetical particle associated with a possible solution to the 
strong CP problem and is a leading candidate for dark matter.  
In this paper 
we investigate the emission of axions by accelerated electrons. We find the emission probability and energy within the WKB approximation for an electron accelerated by an electromagnetic field.
As an application, we estimate the number of axions produced by electrons accelerated using two counter-propagating high-intensity lasers and discuss how they would be
converted to photons to be detected. We find that, under realistic experimental conditions, competitive 
model-independent 
bounds on the coupling between the axion and the electron could be achieved in
such an experiment.   
\end{abstract}

\maketitle


{\it Introduction}.
The Standard Model of Particle Physics is one of the most successful theories of fundamental physics,
whose accuracy has been verified by numerous laboratory experiments.
It has 
some shortcomings, however. 
One of them is that it allows for  
CP violation in the strong sector. The neutron electric dipole moment, which is a natural consequence of this model, 
has been constrained by experiments to be unexpectedly negligible. The Standard Model also fails to 
accommodate the existence of dark matter. 
Peccei and Quinn \cite{Peccei:1977hh,Peccei:1977ur} proposed to
solve the strong CP problem  
by introducing an anomalous $U(1)$
symmetry. 
Later, Weinberg and Wilczek noted the presence of a pseudo-Goldstone boson, the axion, due to the spontaneous breaking of 
this $U(1)$ symmetry \cite{Weinberg:1977ma,Wilczek:1977pj}. This hypothetical
particle, the axion, is also a possible dark matter candidate \cite{Preskill:1982cy,Abbott:1982af,Dine:1982ah}. 

In many of the viable models, the axion, which is a massive particle, 
couples to both the photon and the electron.
So far there is no experimental evidence for an axion,and numerous experiments and astrophysical searches have been used to put limits on the axion parameter space. For example, the CAST experiment \cite{CAST:2017uph} converts axions, if they exist, into photons using a strong magnetic field. These axions are hypothetically produced in the Sun by Primakoff scattering. However, 
the upper bound on the axion-photon coupling in astrophysical searches
can be model-dependent. 
For example, in the case of the Sun, collective plasma screening can change the rate of axion production by Primakoff scattering, and 
hence, the inferred limits suffer from these systematic uncertainties. For this reason, purely terrestrial experiments, where the (hypothetical) production and detection of axions are model-independent, have an important role. 
One such experiment is the Optical Search for QED Vacuum Bifringence, Axions and Photon Regeneration (OSQAR) \cite{OSQAR:2015qdv}, which uses a light-shining-through-walls (LSW) approach to produce and detect axions in the laboratory. Other LSW experiments have been proposed \cite{Beyer:2020dag,Beyer:2021mzq} for which the production of axions is achieved using high-power laser beams. For a comprehensive review of current axion searches, refer to Ref.~\cite{RevModPhys.93.015004} and the references therein. The experiments mentioned above only test the coupling between an axion and two photons, and therefore can put constraints on the photon-axion coupling but not on the electron-axion coupling.
In this paper, we propose a different mechanism for axion production in the laboratory which makes use of the electron-axion coupling. Accelerated electrons can emit axions
in a way similar to the Larmor radiation of photons.
This production mechanism allows us to  
constrain the electron-axion coupling, $g_{ae}$, in a model-independent manner.
Although laboratory-based constraints on $g_{ae}$ are currently available from nuclear reactor experiments \cite{PhysRevLett.124.211804}, the method we propose here has the advantage of being scalable.
Its scalability allows us 
to lower the upper bound on $g_{ae}$ with the progress of laser technology, which is expected to continue in the next decade with numerous new laser facilities already proposed \cite{mp3}.
 

Throughout this paper, metric signature $(+,-,-,-)$ and units where $c= \epsilon_0= k_B=1$ are used unless stated otherwise. Initially, Planck's constant $\hbar$ is not set to 1 because we use a semi-classical expansion in powers of $\hbar$.

\vspace{6pt}

{\it Axion production.} 
We study the emission of axions by an 
electron described by a spinor field $\psi$ which is accelerated through
a classical electromagnetic potential $A_\mu$. 
We use the semi-classical solution, i.e. the solution in the WKB approximation,  
to the Dirac equation in an external electromagnetic field.
Here, we present the main results and leave the details of the calculations to the appendix. We note that photon emission using a semi-classical approximation is well known and has been studied in the context of the Baĭer-Katkov method \cite{BAIER1967492, Baier_Katkov_2, Baier_Katkov_3, Baier_1972}.

The free Lagrangian is given by
\begin{equation}
    \mathcal{L}_{\text{free}} = i \hbar \overline{\psi} \gamma^\mu D_\mu \psi - m \overline{\psi}\psi + \frac{1}{2} \partial_\mu \phi \partial^\mu \phi - \frac{m_a^2}{2 \hbar^2}  \phi^2 \,,
\end{equation}
where $\phi$ is the axion field and and $m_a$ its mass. The electron mass is denoted by $m$ and the covariant derivative is $D_\mu = \partial_\mu  -i e A_\mu / \hbar$, where $-e$ is the electron charge. The presence of the potential modifies the Dirac equation as follows
\begin{equation}
    i \gamma^\mu (\hbar \partial_\mu - i eA_\mu)  \chi - m \chi = 0 \, . 
\label{diraceq}
\end{equation}
We let $\chi=\Psi e^{-iS/\hbar}$
and find the scalar function $S$ and spinor $\Psi$ to zeroth order in $\hbar$.
By applying $[i \gamma^\mu (\hbar \partial_\mu - i eA_\mu)  + m]$ on the left in \cref{diraceq}
we find that $S$ must satisfy
\begin{align}
   (\partial^\mu S+eA^\mu)(\partial_\mu S+eA_\mu) - m^2 = 0 \,.
\label{0thWKB}   
\end{align}

We assume that $A_\mu=0$ on the hypersurface $t=t_i$ in the past. Then, we consider the world lines with a uniform
velocity emanating from this hypersurface towards the
future. We define $\tau$ as the proper time along these world
lines measured from the $t = t_i$ hypersurface and we assume
that these world lines do not cross each other. We let these world lines obey the classical equations of motion for 
a charged particle with charge $-e$,
\begin{align}
    \frac{d^2 x^\mu }{d \tau^2} = - \frac{e}{m} F^{\mu \nu} \frac{d x_\nu}{d \tau} \,,
    \label{lorentz-equation}
\end{align}
where $F_{\mu \nu} = \partial_\mu A_\nu - \partial_\nu A_\mu$. Then, the solution to \cref{0thWKB} exists and is given by
\begin{align}
    \partial_\mu S = m v_\mu - eA_\mu \,, \quad v^\mu = \frac{d x^\mu}{d \tau} \,.
 \label{WKB0thsol}   
\end{align}
In particular, the vector field $m v_\mu - eA_\mu$ remains
hypersurface orthogonal as a consequence of eq.~\eqref{lorentz-equation}.
Then, \cref{diraceq} at zeroth order in $\hbar$ is solved, in the representation of the $\gamma$-matrices
such that $\gamma^0$ is diagonal~\cite{BjorkenDrell}, by 
\begin{align}
 \Psi = \sqrt{p_0 +m} \begin{pmatrix}
     s \\
     \frac{\boldsymbol{\sigma}\cdot \mathbf{p}}{p_0 + m}s
 \end{pmatrix}\exp\left( - \frac{1}{2}\int_0^\tau \partial_\mu v^\mu(\tau')\,d\tau'\right)\, ,
 \label{soliwildtilePsi}
\end{align}
where $p^\mu = m v^\mu$. A condition 
for the first-order correction to $\Psi$ to exist leads to
%
the Thomas-BMT equation \cite{Thomas:1926dy,Bargmann:1959gz} for the two-dimensional spin state $s$:
\begin{align}
\begin{split}
 & \frac{ds}{d\tau}
    = - i\mathbf{F}\cdot\boldsymbol{\sigma}s\, , \quad \mathbf{F} = \frac{e}{2m}\left(\mathbf{B} - \frac{\mathbf{p}\times\mathbf{E}}{p_0+m}\right)\, ,
\end{split} 
\label{spinrotation}
\end{align}
where $\mathbf{E}$ and $\mathbf{\mathbf{B}}$ are the electric and magnetic field, respectively. We have verified 
that \cref{spinrotation} is Lorentz invariant as expected. 

Now, we expand the electron field using a basis consisting of
wave-packet solutions as
\begin{align}
    \psi(x) = \sum_{\mathbf{p},\alpha} \left[ u_{(\mathbf{p},\alpha)}(x) b_{(\mathbf{p},\alpha)} +  v_{(\mathbf{p},\alpha)}(x) d^\dagger_{(\mathbf{p},\alpha)} \right] \,.
\label{elecexp}    
\end{align}
Here, the modes $u_{(\mathbf{p},\alpha)}, v_{(\mathbf{p},\alpha)}$ are wave packets which have  approximately definite world lines and 
four-momentum $\mathbf{p}$ of a classical particle and polarization state $\alpha$,
with positive and negative energy, respectively, 
and satisfy usual orthogonality conditions (see App. II).
The wave-packet modes with different (discrete) momenta are orthogonal to one another.
Then, the annihilation and creation 
operators satisfy the following anti-commutation relations: 
\begin{align}
    \left\{ b_{(\mathbf{p},\alpha)}, b_{(\mathbf{p},\beta)}^\dagger \right\} = \left\{ d_{(\mathbf{p},\alpha)}, d_{(\mathbf{p},\beta)}^\dagger \right\} = \delta_{\alpha \beta} \,.
\label{elecanticomm}    
\end{align}
The operators with different momenta anti-commute. 

We approximate the modes $u_{(\mathbf{p},\alpha)}$ by a superposition of the WKB
solutions we discussed earlier. Then, they take
approximately the following form:
\begin{align}
    u_{(\mathbf{p},\alpha)}(x) = \sqrt{\frac{p_0+m}{2p_0}} \begin{pmatrix}
        s_\alpha \\
        \frac{\mathbf{p}\cdot\boldsymbol{\sigma}}{p_0 + m} s_\alpha 
    \end{pmatrix} G(x) \,,
\label{elecmodeu}
\end{align}
where $s_\alpha$ satisfies \cref{spinrotation} and $G(x)$ is peaked around a classical world line with the tangent vector $v^\mu = p^\mu/m$.  The factor $e^{-i S/\hbar}$ and the exponential factor in 
\cref{soliwildtilePsi} are absorbed into this function, which is normalized as
\begin{align}
    \int d^3 \mathbf{x} \; \left| G(t,\mathbf{x}) \right|^2 =1 \, .
\end{align}
The function $G(x)$ is smooth but $|G(x)|^2$ can be approximated by $\delta^{(3)}(\mathbf{x} - \mathbf{x}(t))$ where $(t,\mathbf{x}(t))$ is the classical world line of the electron. 

The free axion field satisfies the Klein-Gordon equation $(\square + m_a^2 / \hbar^2) \phi=0$ and can be expanded as
\begin{equation}
    \hat{\phi}(x) = \int \frac{d^3 \mathbf{k}}{2k_0 (2 \pi)^3} [a_\mathbf{k} e^{-i k \cdot x } + a^\dagger_\mathbf{k} e^{i k \cdot x }] \,,
\label{axionexp}    
\end{equation}
where $k_0 = \sqrt{k^2 + (m_a/ \hbar)^2}$ and the operators satisfy 
\begin{align}
\left[a_\mathbf{k},a^\dagger_\mathbf{k'}\right] = (2 \pi)^3 2 \hbar k_0 \delta^{(3)}(\mathbf{k}-\mathbf{k'}) \,.
\label{axionopscommrelation}
\end{align}
We approximate the axion field by a massless field, letting $m_a = 0$, for simplicity. Thus, the results obtained here are valid as long as the axion mass is much smaller than the typical axion momentum. As will be shown, we can impose bounds on the axion parameter space for $m_a \lesssim10$ keV, although a massless limit was assumed initially. Notice that the wave number of the axion is regarded as classical so that its
momentum is of order $\hbar$ in our WKB approximation. 
(This is similar to the case for photons~\cite{Higuchi:2005an}.)
The interaction Lagrangian between the axion and the electron field is \cite{Raffelt:1996wa}
\begin{align}
    \mathcal{L}_{\textrm{int}}
    & = -\frac{\hbar g_{ae}}{2m}\partial_\mu \phi \overline{\psi}\gamma_5 \gamma^\mu \psi\,.
\label{intLpseudo}
\end{align}
To leading order in $g_{ae}$, the interaction Hamiltonian is $\mathcal{H}_{\text{int}} = - \mathcal{L}_{\textrm{int}}$ with normal ordering. We assume that the initial electron state $b_{(\mathbf{p},\alpha)}^\dagger \ket{0}$, where $\ket{0}$ is the vacuum state, is such that the final state (after the emission of an axion with typical momentum $\mathbf{k}$) is approximately $a^\dagger_\mathbf{k} b_{(\mathbf{p},\beta)}^\dagger \ket{0}$. Given the large energy difference between the electron and the axion, it seems reasonable to take (approximately) the same initial and final wave-packet states for the electron wavefunctions. A similar approach leads to the model of photon emission from the classical electron starting from QED (see Ref.~\cite{Martin:2007equ}). The only possible difference between the initial and final state is the spin state. Thus, 
the electron spin is either flipped or not flipped after emission, but the spatial
wave function remains the same in our approximation.

To first order in perturbation theory, the total one-axion-emission 
final state is given by
\begin{align}
\begin{split}
\ket{f}  = -\frac{i}{\hbar} \int d^4 x \; \mathcal{H}_{\text{int}}(x) b_{(\mathbf{p},\alpha)}^\dagger \ket{0} \,,
\label{finalstate}
\end{split}
\end{align}
and the emission probability is $P_{\text{em}}= \braket{f|f}$. Inserting \cref{elecexp,axionexp} into \cref{finalstate} and using the operator relations~\eqref{elecanticomm} and 
\eqref{axionopscommrelation}, we find
\begin{align}
P^{(\beta,\alpha)}_{\textrm{em}} = \hbar \int \frac{d^3\mathbf{k}}{(2\pi)^32k_0}
|\mathcal{A}_{(\mathbf{p},\mathbf{k},\beta,\alpha)}|^2\,. \label{eq:general-emission-probability}
\end{align}
The interaction amplitude is 
\begin{align}
    \mathcal{A}_{(\mathbf{p},\mathbf{k},\beta,\alpha)} = \frac{g_{ae}}{2m}\int d\tau\, \frac{e^{ik\cdot x(\tau)}}{(k\cdot v)^2}
    s_\beta^\dagger(\tau)\mathbf{Q}(\tau)\cdot\boldsymbol{\sigma}s_\alpha(\tau)\,,
    \label{the-amplitude2}
\end{align}
where 
\begin{align}
\begin{split}
 \mathbf{Q} & = \mathbf{V}
    - (k\cdot a)\mathbf{k}
    - \left[V_0 - (k\cdot a)k_0\right]\frac{\mathbf{p}}{p_0+m}\;,   \\
    V^\mu & = \frac{e}{m}(k\cdot v)F^{\mu\nu}k_\nu\,.
\label{QandVdef}    
\end{split}    
\end{align}
Here, $a^\mu = d v^\mu / d \tau$ is the proper acceleration of the electron. The spin states $s_{\alpha}$ and $s_{\beta}$ satisfy \cref{spinrotation}.
We verify \cref{eq:general-emission-probability} to lowest order in $eA_\mu$ by using a Feynman-diagram calculation in App. III. 

In the next section, we study laser-accelerated electrons, which follow two-dimensional trajectories. 
In this case, we choose an electric field in the $xz$-plane and a magnetic field parallel to the $y$-axis. It is convenient to choose the initial spin polarization along the $y$-direction. Then, 
\begin{align}\label{2D-nf-f}
 \begin{split}
 P_{\mathrm{em}}^{\mathrm{nf}} & =\frac{\hbar g_{ae}^2}{4m^2}
    \int \frac{d^3\mathbf{k}}{(2\pi)^32k_0}k_y^2
    \left|\int d\tau\,e^{ik\cdot x}\frac{k\cdot a}{(k\cdot v)^2}\right|^2\,, \\
    P_{\mathrm{em}}^{\mathrm{f}} & = \frac{\hbar g_{ae}^2}{4m^2}
    \int \frac{d^3\mathbf{k}}{(2\pi)^32k_0}
    \left| \int d\tau\, e^{ik\cdot x \mp if(\tau)}[Q_z \pm i Q_x] \right|^2 \,,
 \end{split}   
\end{align}
where $d f(\tau) / d \tau = F_y$, with $F_y$ defined in \cref{spinrotation}, and $Q_z \pm i Q_x$ is found from \cref{QandVdef}. The first expression is the emission probability when the electron does not flip its spin ($\alpha = \beta$) whereas the second corresponds to the spin-flip case.
In the second equation in \eqref{2D-nf-f}, the upper and lower signs are for
the spin states $s_+$ and $s_{-}$, respectively, where
$\sigma_y s_\pm = \pm s_\pm$. 

It is possible to find a simpler expression for the spin-averaged axion energy emitted 
for a three-dimensional motion using the technique 
employed in deriving the Larmor formula in the WKB approximation~\cite{Higuchi:2005an,Martin:2007equ,Higuchi:2009ms}. 
With the definition $n^\mu = k^\mu /k_0$, we find the spin-averaged axion energy emitted as
\begin{align}
\braket{E} & =\frac{\hbar^2 g_{ae}^2}{16\pi m^2}
    \int \frac{d\Omega}{4\pi}
    \int \frac{d\tau}{(n\cdot v)^7} \notag \\
& \times    \left[ -\frac{9(n\cdot a)^2}{(n\cdot v)^2}U_\mu U^\mu  
    + \frac{3n\cdot a}{n\cdot v}\frac{d\ }{d\tau}(U_\mu U^\mu)
    - H_\mu H^\mu\right]\,, \label{eq:starting-point}
    \end{align}
where 
\begin{align}
 \begin{split}
   U^\mu  & = (n\cdot a)n^\mu - \frac{e}{m}(n\cdot v) F^{\mu\nu}n_\nu\,, \\
    H^\mu  & = \frac{dU^\mu}{d\tau} + \frac{e}{m}F^{\mu\nu}U_\nu  \,.
 \end{split}   
\end{align}

\begin{figure}[H]
\centering 
\includegraphics[width=1\linewidth]{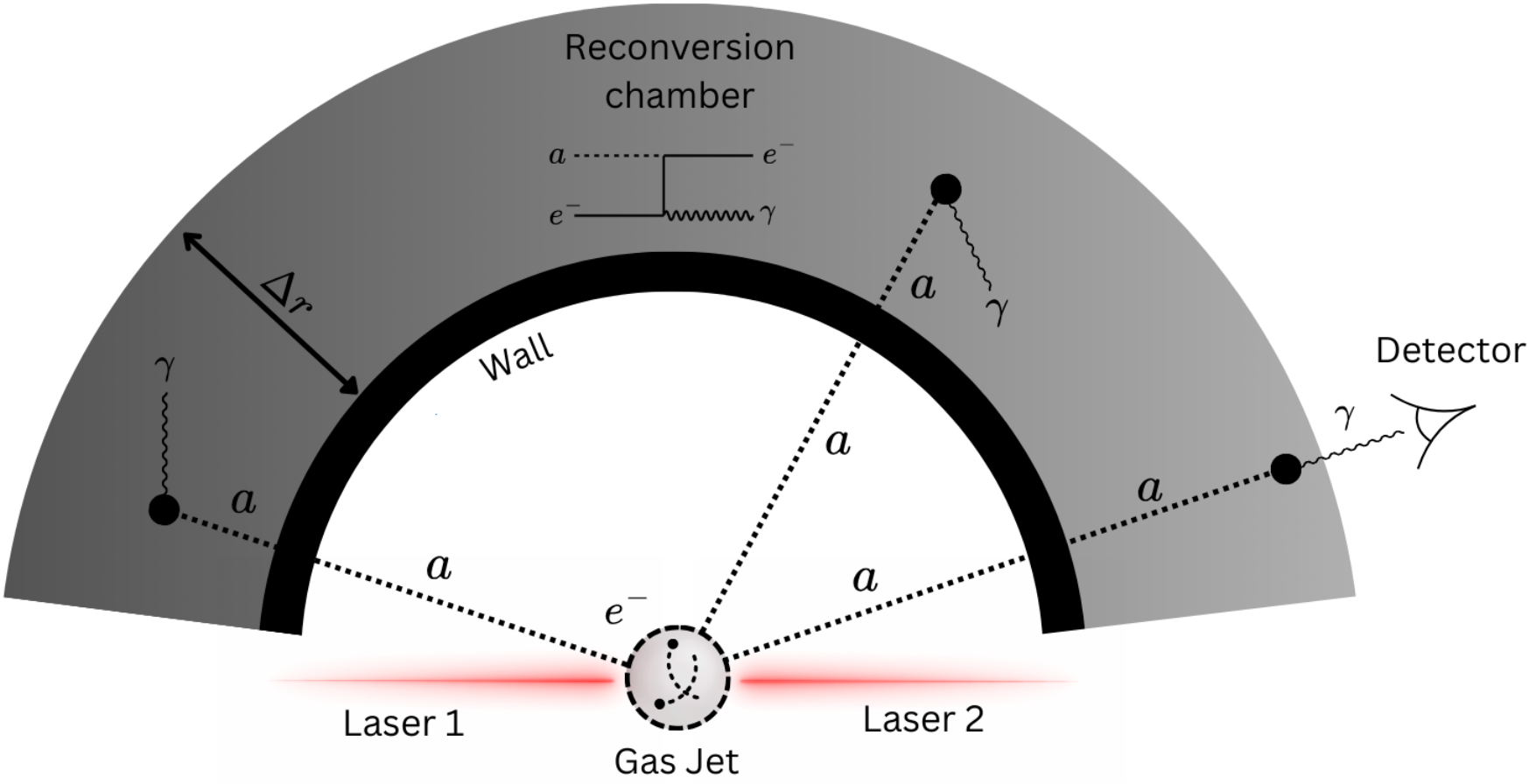}
\caption{Diagram of the experimental proposal. We assume that there are enough detectors to cover a large solid angle.}
\label{fig1}
\end{figure}

\vspace{6pt}
{\it Experimental proposal}.
The proposed setup consists of an emission region where the axions are produced by laser-accelerated electrons and a conversion region where they are re-generated into photons (see \cref{fig1}).
Throughout this section, we put $\hbar =1$ and assume that the parameters $m_a$, $g_{ae}$ and $g_{a\gamma}$ are independent. For the results of the previous section to be applicable, $m_a \ll k_a$ where $k_a$ is the typical axion momentum, 
which 
depends on the laser parameters. 
We have $k_a \sim 10^2$\,keV for 
the laser parameters we will consider.  Therefore, the bounds on $g_{ae}$ obtained in this section will be valid for $m_a \lesssim 10$ keV. The axions are converted to photons through processes $a +e^{-} \to \gamma+e^{-}$ and $a + N_T \to \gamma +N_T$ that involve the couplings $g_{ae}$ and $g_{a\gamma}$,  respectively ($N_T$ here is an atomic target) \cite{PhysRevD.34.1326,PhysRevLett.123.071801}. Only the former process is considered as it allows us to impose upper bounds on $g_{ae}$ in the absence of photon detection. In what follows, we use the model-independent bound $g_{a \gamma} \lesssim10^{-4}$GeV$^{-1}$ \cite{PhysRevLett.124.211804}. 

A hydrogen gas jet is used as a source of electrons.
These are accelerated by two counter-propagating lasers of pulse duration $\tau_p$, forming a standing laser field as described in \cite{PhysRevLett.83.256}. 
The beams are linearly polarized along the $z$-direction and propagating along the 
$x$-direction. They are described as plane waves with angular frequency $\omega_0$. 
The electric and magnetic fields are therefore
\begin{equation}
\begin{aligned}& E_z = E_0  \left[ \cos {\omega_0 (t - x)} + \cos {\omega_0 (t + x)}\right] \, ,  \\ 
& B_y = - E_0  \left[ \cos {\omega_0 (t - x)} -\cos {\omega_0 (t + x)} \right] \, .  
\end{aligned}
\end{equation}
The classical equation of motion of the electron is
\begin{equation}
\begin{aligned}
  \frac{d p_x}{d t} =  e \beta_z B_y \,, \quad \frac{d p_z}{d t } = -e(E_z + \beta_x B_y) \,,
\end{aligned}  \label{eomelectron}  
\end{equation}
where $(\beta_x,\beta_y,\beta_z)$ is the velocity divided by the speed of light. Electrons placed at the nodes $ \omega_0 x = 2\pi n, n \in \mathbb{Z}$ do not feel the effects of the magnetic field and have unstable oscillatory trajectories, which implies that it is difficult to realize them experimentally. 
The trajectories of the off-node electrons, which constitute the vast majority of the electrons accelerated by the laser beams, need to be found by solving \cref{eomelectron} numerically.
In principle, it would be possible to use \cref{2D-nf-f} with these trajectories and estimate the number
of axions emitted numerically.  However, such computations would be quite challenging.
Instead, we proceed as follows.
We first find the energy emitted in the presence of the magnetic, as well as electric, field by \cref{eq:starting-point}. We then estimate the number of emitted axions by dividing the total energy by the typical axion energy, which we assume to be much larger than $m_a$. This is justified if the spectrum is peaked around the typical energy. For an electron in a constant magnetic field we find that the spectrum is peaked and the typical energy of individual axions is around $4 a_0^3 \omega_0$ (see App. V), where $a_0 = e E_0/m \omega_0$ is the laser strength parameter. Although for laser beams the magnetic field is not constant, we assume the axion spectra to be similar.
We note that, since for relativistic electrons the axions are predominantly emitted along the trajectory, the axion emission is concentrated in the $xz$-plane. 

We find that the WKB treatment of the previous section is valid if $ a_0^2 \omega_0 /m \ll1$ or $(a_0/700)^2 \ll1$ for $\omega_0 \approx1$ eV (see App. I).

The energy emitted by an electron in one cycle of duration $2 \pi/ \omega_0$ can be written using \cref{eq:starting-point} as $ \langle E \rangle_{(2\text{d})} = g_{ae}^2 \omega_0^3m^{-2} \mathcal{N}(a_0)$, where $\mathcal{N}(a_0)$ is a number found numerically. 
The total number of electron ``oscillations" is given by $\rho_e V \nu \tau_p n_s/2$ where $\rho_e$ is the electron density, $n_s$ is the number of shots, $V$ is the volume occupied by one beam and $\nu=\omega_0/2\pi$ is
the frequency. We note that $V= A \tau_p$ where $A$ is the beam cross section. Then, the total number of axions produced is 
\begin{equation}
    N_a^{tot} = \frac{g_{ae}^2 \rho_e V \tau_p \omega_0^3 n_s \mathcal{N}(a_0)}{16 \pi a_0^3 m^2} \,,
\label{totalnbaxions}    
\end{equation}
which was obtained by dividing the total energy by the typical axion energy 
$k_a = 4 a_0^3 \omega_0$. 

As shown in \cref{fig1} (and as in many LSW experiments), axions produced by accelerated electrons pass through a wall which prevents
background photons, 
e.g., those produced by Larmor radiation, from entering the detector. We note that the effect of Larmor radiation on the trajectories can be neglected if $(a_0/250)^2\ll1$.
The reconversion occurs because the axions then interact with the electrons in the material that surrounds the wall. 
The Compton-like process $a + e^{-} \to \gamma + e^{-}$ has total cross section \cite{Avignone:1988bv}
\begin{equation}
    \sigma^c = \frac{Z \alpha g_{ae}^2}{4 m^2} f\left( \frac{k_a}{m} \right), \quad f(x) = \frac{1}{x} \ln{(1+2x)} - \frac{2(1+3x)}{(1+2x)^2}\,,
\end{equation}
in the limit $m_a \to 0$, where $\alpha$ and $Z$ are the fine structure constant and the atomic number, respectively. As before, we let $k_a = 4 a_0^3 \omega_0$. The photon energy is $\sim k_a$ for $m \gg k_a \gg m_a$. The total number of produced photons is
\begin{equation}
    N_\gamma = \sigma^c \rho_m \Delta r N_a^{tot} \, ,
\label{nbphotons}    
\end{equation}
where $\rho_m$ is the atom density of the material used for the reconversion and $\Delta r $ is its length. The latter is at most of the order of the photon attenuation length. Not all produced photons will enter the detector, as the ones emitted backwards will be absorbed by the wall. However, this effect will only result in a $\mathcal{O}(1)$ correction factor in \cref{nbphotons}. Substitution of \cref{totalnbaxions} into \cref{nbphotons} gives
\begin{equation}
    N_\gamma = \frac{Z \alpha^2 g_{ae}^4}{8 a_0^5} \mathcal{N}(a_0) f \left( \frac{4 a_0^3 \omega_0}{m}\right) n_s \frac{\rho_m \rho_e \tau_p \Delta r \omega_0 E_{las}}{m^6} \, ,
\end{equation}
where $E_{las} = A \tau_p m^2 \omega_0^2 a_0^2/(8 \pi \alpha)$ is the energy of the laser pulse. 

Let us estimate the bound on $g_{ae}$ achievable for a laser with $E_{las} =1\,$kJ, wavelength $\lambda \approx 1$  $\mu\,$m, $\tau_p =10^{-12}\,$s, beam diameter $\sim 10 \lambda$ and a repetition rate of $10\,$Hz. Such a laser is not currently available, but is within the current technological capabilities. For this choice $a_0 \approx 30$, and $E_\gamma \approx k_a \approx 133\,$keV. We find numerically that $\mathcal{N}(30) \approx 10^{13}$. We also assume that the source of electrons is a hydrogen gas for which $\rho_e = 10^{20}\,$cm$^{-3}$. For the given density and intensities, plasma effects can be neglected. For reconversion, we use aluminum. The attenuation length for these photon energies is $\Delta r \sim 1\,$cm. If $N_\gamma \lesssim 1$ after one week of measurement, we find that $g_{ae} \lesssim 4.1 \times 10^{-5}$. This result is valid for $m_a \ll k_a (\sim 10^2\,$keV). Next-generation lasers could reach $E_{las} = 10^2\,$kJ, $\tau_p = 10^{-10}$ s with a repetition rate of $1\,$MHz. 
Given the rapid development of diode laser technology, it is not inconceivable that such a laser would become available in the future. Assuming that the remaining parameters stay the same, if $N_\gamma \lesssim 1$ after one year of measurement, we find $g_{ae} \lesssim 8.5 \times 10^{-8}$.  
\newline
The axions can also decay to photons with a decay rate $\Gamma(a \to \gamma \gamma) = g_{a\gamma}^2 m_a^3 / 64 \pi$. The survival probability of the axion to reach the reconversion chamber at distance $\ell$ is $P_{sur} = \exp[- \ell m_a \Gamma / k_a]$ \cite{PhysRevLett.124.211804}. However, for $\ell \sim 1\,$m, $m_a \lesssim10$ keV and $g_{a\gamma} \lesssim 10^{-4}$ GeV$^{-1}$ \cite{PhysRevLett.124.211804}, $P_{sur} \approx 1$. 

\begin{figure}[h]
	\centering
	\begin{tikzpicture} 
		\begin{groupplot}[group style={group size=1 by 2,horizontal sep=0.65cm, vertical sep=0.5cm},xmin=0,ymin=0,width=15cm]
			\nextgroupplot
			[
			ylabel={$g_{ae}$},
                xlabel={$m_a$ (eV)},
			enlargelimits=false,
			axis on top,
			xmin=1e-1, xmax=1e5,
			ymin=1.5e-10, ymax=3e-2,
			width=0.96\linewidth,
			height=0.72\linewidth,
                xmode=log,
                ymode=log,
                xtick={0.1, 1, 10, 100, 1000, 10000, 100000},
                xticklabels={$10^{-1}$, $10^{0}$, $10^{1}$, $10^{2}$, $10^{3}$, $10^{4}$, $10^{5}$,}
			]
			
			\addplot graphics [xmin=1e-1,xmax=1e5,ymin=1.5e-10,ymax=3e-2] {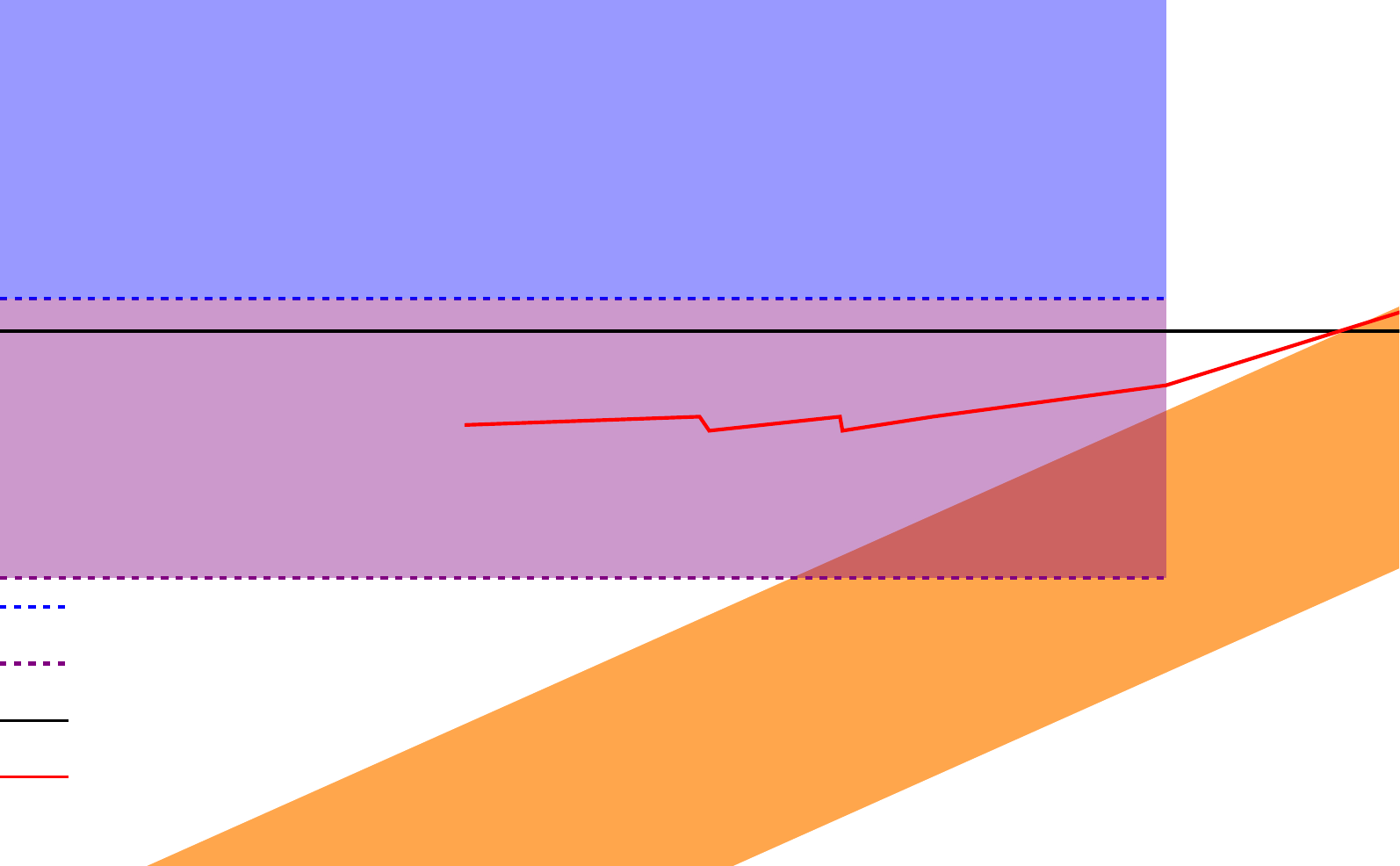};
			
            \node[anchor= south west] at (rel axis cs:0.04, 0.254) {\footnotesize Current lasers};
		\node[anchor= south west] at (rel axis cs:0.04, 0.19) {\footnotesize Future lasers};
            \node[anchor= south west] at (rel axis cs:0.04, 0.115) {\footnotesize Electron g-2};
            \node[anchor= south west] at (rel axis cs:0.04, 0.06) {\footnotesize Reactor};

            \node[anchor= south west] at (rel axis cs:0.5, 0.15) {\footnotesize DFSZ};
			
		\end{groupplot}
	\end{tikzpicture}
	\caption{Bounds of laboratory-based experiments. The red and black curves correspond to the bounds found in \cite{PhysRevLett.124.211804}, and \cite{Constraining_exotic_spin} respectively. The orange band is the DFSZ prediction \cite{Zhitnitsky:1980tq, Dine:1981rt, PhysRevD.33.897}. For the next-generation laser, we assumed one year of measurement instead of one week.}
\label{fig3}
\end{figure}
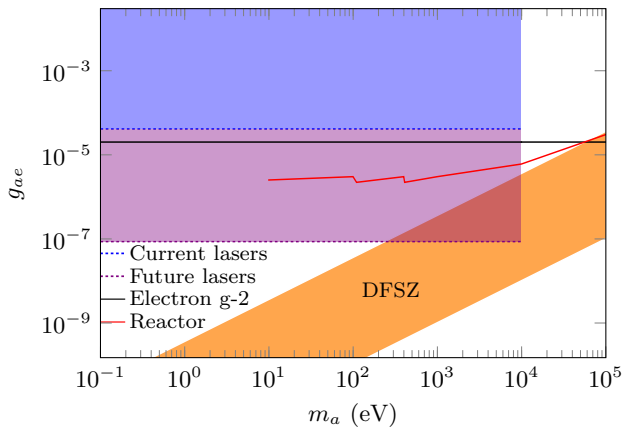

Axion searches have been conducted in the range of masses $m_a \lesssim 10\,$ keV. The experiments XENON1T \cite{PhysRevD.104.023019} and XENONnT \cite{PhysRevLett.129.161805} were used to impose stringent bounds on $g_{ae}$ for solar axions. Constraints on $g_{ae}$ for laboratory-produced axions were found using anomalous magnetic moment and electric dipole moment \cite{Constraining_exotic_spin} or nuclear reactors for neutrino experiments \cite{PhysRevLett.124.211804}. These bounds are shown in \cref{fig3}. In the same figure, we also report the projected bounds from the proposed laser experiment. Using current laser technology and just a week of data gathering, we expect to reach exclusion bounds comparable to other laboratory searches. On the other hand, with a future laser system we could expect to achieve bounds up to the prediction of DFSZ axions \cite{Zhitnitsky:1980tq, Dine:1981rt, PhysRevD.33.897}. The DFSZ axion model followed that of Weinberg and Wilczek, as well as KSVZ \cite{PhysRevLett.43.103, Shifman:1979if} and predicts a coupling of the axion to electrons and light quarks with the coupling constant being proportional to the axion mass.

\vspace{6pt}

{\it Conclusion.} In this paper, we used the WKB approximation to calculate the number of axions and the energy emitted from an electron accelerated in an electromagnetic field. We then applied the WKB approximation to the case of a two-dimensional electron trajectory, where the electron was accelerated by laser beams. Applying the results to an experimental proposal allowed us to find bounds on the coupling $g_{ae}$. The bounds found are valid for axion masses $m_a \lesssim 10\,$keV. Since the results derived in the first section are general, they can be used for other experimental proposals (e.g. replacing the linearly beams by circularly polarized ones).
We compared the bounds obtained here with those obtained in other laboratory experiments (e.g. axion production achieved using a nuclear reactor) and found that they are comparable. Whereas it is unlikely that the performance of nuclear reactors will be largely improved in the near future, lasers are subject to important performance enhancements and therefore could allow  
further exploration of the parameter space of the QCD-axion.  

\vspace{6pt}
{\it Note added.} Recently, we have become aware of Ref.~\cite{PhysRevD.103.076011} which also derives the WKB electron wave function in a form similar to the one given in this paper.  
\vspace{6pt}

{\it Acknowledgments.} G.V. acknowledges helpful discussions with Dr. Konstantin A. Beyer. This work was supported in part by EPSRC Grants No. EP/X01133X/1 and No. EP/X010791/1. G. G. is also a member of the Quantum Sensing for the Hidden Sector(QSHS) Collaboration, supported by STFC Grant No. ST/T006277/1. 
\newline
{\it Data availability.} The code used to find the numerical results is available in Ref.\cite{code}.

\onecolumngrid
\section*{End Matter}
\begin{figure}[H]
    \centering
    \begin{tikzpicture}
        \begin{groupplot}[
                group style={
                    group size=1 by 1,
                    vertical sep=0cm,
                    horizontal sep=0cm
                    },
                xmin=5,
                ymin=0,
                height={0.72*\linewidth},
                width=0.96\linewidth,
                no markers
                ]
                
            \nextgroupplot
            [
            xlabel={$a_0$},
            enlargelimits=false,
			axis on top,
            ylabel={$\frac{\langle E \rangle}{k_a} \, \left( \frac{g_{ae}^2 \omega_0^2}{m^2}\right)$},
            xmin=5, xmax=50,
            ymin=1, ymax=10000000000,
            ymode=log,
            tick align=inside,
            tick pos=both,
         scaled ticks = false,
            scaled y ticks = false,
            label style={font=\small},
            tick label style={font=\small},
            ]

            \addplot[
			color=blue, very thick, dashed] table [col sep = comma, x index = 0, y index = 1]{on-node-Na.txt};

            \addplot[
			color=red, very thick] table [col sep = comma, x index = 0, y index = 1]{off-node-Na.txt};

        \end{groupplot}
    \end{tikzpicture}
    \caption{Number of axions produced in one cycle (estimated by dividing the total energy by the typical axion energy) by one on-node electron (blue-dashed line) and by one off-node electron (red-solid line) with initial condition $\omega_0 x(0) = \pi/3$. The number was averaged over ten cycles.}
    \label{fig2}
\end{figure}
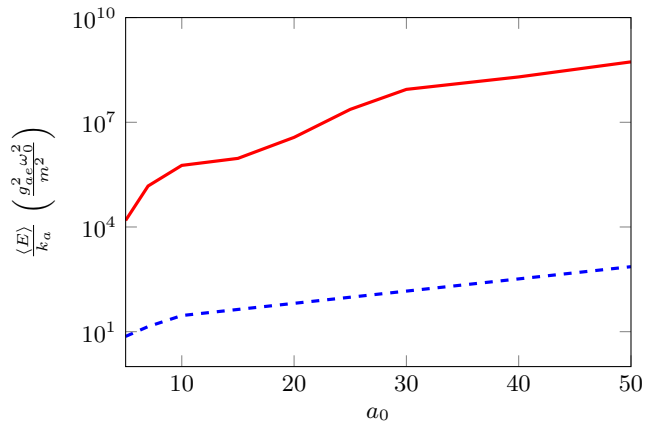

\twocolumngrid
\textit{Appendix: Comparison between on- and off-node electrons}. We show here the effect of the magnetic field on axion production. On-node electrons have an oscillatory trajectory with
\begin{align}
    \gamma \beta_z = - 2a_0 \sin{\omega_0 t}, \quad \gamma= \sqrt{1 + 4 a_0^2 \sin^2{\omega_0 t}} \, .
\end{align}
The energy emitted in one cycle is given by
\begin{align}
    \langle E \rangle_{(1\rm{d})} = \frac{g_{ae}^2 \omega_0^3}{6 m^2} (7a_0^4 + a_0^2) .
\end{align}
The typical axion energy for on-node electrons is $4  a_0^2 \omega_0$ and the spectrum is peaked around this value. The particle numbers emitted in one cycle by on- and off-node electrons are shown in \cref{fig2}. The number of emitted axions is much larger for the off-node case, which shows the importance of the magnetic field on particle production.

\appendix

\section*{Appendix I: WKB solutions with general electromagnetic field}
\label{appI}
The Dirac equation in an external electromagnetic 
potential $A_\mu$ is
\begin{align}
    i\gamma^\mu(\hbar\partial_\mu - ieA_\mu)\psi - m\psi = 0\,. \label{eq:original-Dirac-eq}
\end{align}
The solutions to this equation in the WKB approximation for a purely time-dependent potential was found in Ref.~\cite{Martin:2007equ}.  We seek an approximate solution
in the form
\begin{align}
    \psi = \check{\Psi} \exp\left(-\frac{i}{\hbar}S\right)\,,  \label{eq:WKB-ansatz-2}
\end{align}
where $S$ is a scalar function.  Substituting eq.~\eqref{eq:WKB-ansatz-2} into
eq.~\eqref{eq:original-Dirac-eq}, one finds
\begin{align}
    [\gamma^\mu(\partial_\mu S + eA_\mu) - m]\check{\Psi} + i\hbar\gamma^\mu \partial_\mu \check{\Psi} = 0\,. \label{eq:first-h-equation}
\end{align}
Letting $\check{\Psi} = \Psi + \Psi^{(h)}$, where $\Psi$ and $\Psi^{(h)}$ are of
zeroth order and of higher order, respectively, in $\hbar$, we can write eq.~\eqref{eq:first-h-equation} \textit{to first order} in $\hbar$ as follows:
\begin{align}
     & [\gamma^\mu(\partial_\mu S + eA_\mu) - m]\Psi = 0\,, \label{eq:WKB-Psi}\\
     & [\gamma^\mu(\partial_\mu S + eA_\mu)- m]\Psi^{(h)}
     + i\hbar\gamma^\mu \partial_\mu \Psi = 0\,. \label{eq:WKB-correction-term}
\end{align}
By multiplying these equations by $\gamma^\mu(\partial_\mu S + eA_\mu) + m$ and
using $\{\gamma^\mu,\gamma^\nu\} = 2g^{\mu\nu}$, we find
\begin{align}
   (\partial^\mu S+eA^\mu)(\partial_\mu S+eA_\mu) - m^2 & = 0\,,
    \label{eq:zeroth-order-WKB-2} \\
   [\gamma^\mu(\partial_\mu S + eA_\mu) + m]\gamma^\nu \partial_\nu \Psi & = 0\,.
   \label{eq:consistency-2}
\end{align}
The WKB solution to the Dirac equation~\eqref{eq:original-Dirac-eq} to lowest order in $\hbar$ is
\begin{align}
    \psi^{(\mathrm{WKB})} = \Psi\exp\left( - \frac{i}{\hbar}S\right)\,, 
\end{align}
where the scalar function $S$ and the spinor $\Psi$ satisfy eqs.~\eqref{eq:WKB-Psi}, \eqref{eq:zeroth-order-WKB-2}
and \eqref{eq:consistency-2}.

Let us find the scalar function $S$ satisfying eq.~\eqref{eq:zeroth-order-WKB-2}.
We first describe the congruence of classical world lines used to 
construct the 
function $S$. These world lines satisfy the Lorentz-force equation,
\begin{align}
    \frac{d^2x^\mu}{d\tau^2} = - \frac{e}{m}F^{\mu\nu}\frac{dx_\nu}{d\tau}\,,
    \label{eq:Lorentz-force-eq-2}
\end{align}
where $\tau$ is the proper time along these world lines.
We assume $A_\mu =0$ on the hypersurface $t=t_i$ in the past where $v^\mu$ is
constant. (Thus, the world lines are parallel to each other when they emanate, with $\tau=0$, from
this hypersurface.)
We assume that these world lines do not cross each other. If 
\begin{align}
    \partial_\mu S + eA_\mu = mv_\mu \,, \label{eq:def-of-v}
\end{align}
where $v^\mu = d x^\mu / d \tau$, then \cref{eq:zeroth-order-WKB-2} is satisfied because $v^\mu v_\mu = 1$. Equation~\eqref{eq:def-of-v} 
can be solved for $S$ if and only if
\begin{align}
    \partial_\mu(eA_\nu - mv_\nu) - \partial_\nu(eA_\mu-mv_\mu) = 0\,,
\end{align}
i.e.,
\begin{align}
    \partial_\mu v_\nu - \partial_\nu v_\mu - \frac{e}{m}F_{\mu\nu} = 0\,,
\label{eq:consistency-for-S-2}
\end{align}
where $F_{\mu\nu} = \partial_\mu A_\nu - \partial_\nu A_\mu$.
Observe that \cref{eq:Lorentz-force-eq-2} follows from this equation by
contracting it by $v^\nu$:
\begin{align}
    v^\nu\partial_\mu v_\nu - v^\nu \partial_\nu v_\mu - \frac{e}{m}F_{\mu\nu}v^\nu=0\,. \label{eq:what-should-be-0}
\end{align}
Since $v^\nu\partial_\mu  v_\nu = 0$
and $v^\nu \partial_\nu = d/d\tau$, we find
\cref{eq:Lorentz-force-eq-2}. 

First we note that \cref{eq:consistency-for-S-2} is satisfied at $t=t_i$,
i.e., at $\tau=0$, because $F_{\mu\nu} = 0$ and $v^\mu$ is constant there. 
Now, the $\tau$-derivative of the left-hand side of \cref{eq:what-should-be-0}
is found, by using \cref{eq:Lorentz-force-eq-2} and the Bianchi identity for
$F_{\mu\nu}$, as
\begin{align}
     & \frac{d\ }{d\tau}\left(
    \partial_\mu v_\nu - \partial_\nu v_\mu - \frac{e}{m}F_{\mu\nu}\right) \notag \\
    & = (\partial_\mu v^\lambda)\left( \partial_\nu v_\lambda - \partial_\lambda v_\nu- \frac{e}{m}F_{\nu\lambda}\right) \notag \\
    & \quad - (\partial_\nu v^\lambda)\left(\partial_\mu v_\lambda
    - \partial_\lambda v_\mu- \frac{e}{m}F_{\mu\lambda}\right)\,.
\end{align}
The unique solution to this equation with the initial condition~\eqref{eq:consistency-for-S-2} at $\tau=0$ is \cref{eq:consistency-for-S-2} for all $\tau$. 
Thus, we have constructed the scalar function $S$ satisfying \eqref{eq:zeroth-order-WKB-2}.  Then, if we define $p^\mu = mv^\mu$, \cref{eq:WKB-Psi} becomes
\begin{align}
    (\gamma^\mu p_\mu  - m)\Psi = 0\,,
\end{align}
and the solution is
\begin{align}
    \Psi \propto \begin{pmatrix} s \\ \frac{\boldsymbol{\sigma}\cdot\mathbf{p}}{p_0+m}s\end{pmatrix}\,, \label{eq:first-order-solution-WKB-2}
\end{align}
where $s$ is a two-component spinor, as is well known. 

The differential equation describing the time evolution of $\Psi$ can be found
from \cref{eq:consistency-2} as follows. 
This equation can be written, using \cref{eq:def-of-v} as
\begin{align}
    (\gamma^\mu v_\mu + 1)\gamma^\nu\partial_\nu \Psi = 0\,.
\end{align}
We add this equation and
\begin{align}
    \gamma^\mu\partial_\mu [ (\gamma^\nu v_\nu -1)\Psi ] = 0\,,
\end{align}
which follows immediately from \cref{eq:WKB-Psi}, to find
\begin{align}\label{eq:tau-derivative-of-Psi}
    \frac{d\Psi}{d\tau} + \frac{1}{2}\left(\partial_\mu v^\mu + \frac{e}{2m}\gamma^{\mu\nu}F_{\mu\nu}\right)\Psi = 0\,,
\end{align}
where we used \cref{eq:consistency-for-S-2}.

Let
\begin{align}
    \Psi = \widetilde{\Psi}\exp\left( - \frac{1}{2}\int_{0}^\tau \partial_\mu v^\mu(\xi)\,d\xi\right)\,.
\end{align}
Then,
\begin{align}
    \frac{d\widetilde{\Psi}}{d\tau} = - \frac{e}{4m}\gamma^{\mu\nu}F_{\mu\nu}\widetilde{\Psi}\,. \label{eq:Psi-time-evolution}
\end{align}
By recalling that
\begin{equation}
\begin{aligned}\label{eq:F-in-terms-of-EB}
    F_{0i} & = E_i\,,\\
    F_{ij} & = - \frac{1}{2}\epsilon_{ijk}B_k\,,
\end{aligned}
\end{equation}
we find
\begin{align}
    -\frac{e}{4m}\gamma^{\mu\nu}F_{\mu\nu}
    & = - \frac{e}{2m}\begin{pmatrix}  i\boldsymbol{\sigma}\cdot\mathbf{B}
    & \boldsymbol{\sigma}\cdot \mathbf{E} \\
    \boldsymbol{\sigma}\cdot\mathbf{E} & i \boldsymbol{\sigma}\cdot\mathbf{B}
    \end{pmatrix}\,.    \label{eq:gamma-F-in-EB}
\end{align}
We also note that the Lorentz-force equation~\cref{eq:Lorentz-force-eq-2} can be
written in terms of the electric and magnetic fields as
\begin{equation}
\begin{aligned}\label{eq:Lorentz-force-EB-1}
    \frac{dp_0}{d\tau} & = - \frac{e}{m}\mathbf{p}\cdot\mathbf{E}\,,
    \\
    \frac{d\mathbf{p}}{d\tau} & = -\frac{e}{m}(p_0\mathbf{E}+\mathbf{p}\times\mathbf{B})\,. 
\end{aligned}
\end{equation}
Let
\begin{align}
    \widetilde{\Psi} & = \begin{pmatrix} \sqrt{p_0+m}\,s \\ \frac{\boldsymbol{\sigma}\cdot\mathbf{p}}{\sqrt{p_0+m}}s\end{pmatrix}\,,
\end{align}
which satisfies \cref{eq:first-order-solution-WKB-2}.
Then, by using eqs.~\eqref{eq:gamma-F-in-EB} and \eqref{eq:Lorentz-force-EB-1},
we find that \cref{eq:Psi-time-evolution} is satisfied
if
\begin{align}
    \frac{ds}{d\tau} & = - \frac{ie}{2m}\left(\mathbf{B}(\tau) - \frac{\mathbf{p}(\tau)\times\mathbf{E}(\tau)}{p_0(\tau)+m}\right)\cdot\boldsymbol{\sigma}\,s\,.
    \label{eq:Thomas-BMT}
\end{align}
This is the Thomas-BMT equation for relativistic spin dynamics~\cite{thomas1926motion,Bargmann:1959gz}. 
Thus, the WKB solution to first order can be written as
\begin{align}\label{eq:WKB-final-form}
    \psi^{(\mathrm{WKB})} & = \sqrt{p_0+m}\begin{pmatrix} s \\ \frac{\mathbf{p}\cdot\boldsymbol{\sigma}}{p_0+m}\,s\end{pmatrix}\exp\left( - \frac{1}{2}\int_0^\tau \partial_\mu v^\mu(\xi)\,d\xi\right)
    \notag \\
    & \quad \times
    \exp\left( - \frac{i}{\hbar}S\right)\,,
\end{align}
where $S$ satisfies \cref{eq:def-of-v}, $p_\mu = m v_\mu$ and $s$ satisfies \cref{eq:Thomas-BMT}, whose solution is given by
\begin{align}
    s_\alpha(\tau) & = T\exp
    \left[ - i\int_{0}^{\tau}\mathbf{F}(\xi)
\cdot\boldsymbol{\sigma}\,
d\xi\right]s_\alpha(0)\,,
\end{align}
where $T$ denotes time-ordering and where we defined
\begin{align}
    \mathbf{F} = \frac{e}{2m}\left(\mathbf{B} - \frac{\mathbf{p}\times\mathbf{E}}{p_0+m}\right)\,. \label{eq:definition-of-F}
\end{align}
It is interesting to note that
\begin{align}
    \overline{\psi}\gamma^\mu\psi = 2mv^\mu \exp\left( - \int_0^\tau\partial_\mu v^\mu(\xi)\,d\xi\right)\,,
\end{align}
and that the current conservation equation
$ \partial_\mu(\overline{\psi}\gamma^\mu\psi)=0$ is satisfied.

The WKB approximation would break down if the congruence of world lines of the classical electron 
had a focal point where $\partial_\mu v^\mu \to -\infty$ as
can be seen in Eq.~\eqref{eq:WKB-final-form}.  We have observed numerically that these world lines tend not to have focal points in the background electromagnetic field we assume in this paper though we cannot exclude them.  Even if focal points occur, we expect that the WKB approximation will be valid away from them as we argue below.

The WKB approximation is expected to be accurate if the first-order 
correction $\Psi^{(h)}$ to $\Psi$ is much smaller than $\Psi$.  It can be found
from Eq.~\eqref{eq:WKB-correction-term} as
\begin{align}
   \Psi^{(h)} = \frac{i\hbar}{2m}\gamma^\mu \partial_\mu \Psi\,,
\end{align}
with the help of Eq.~\eqref{eq:consistency-2}.  Hence, the WKB approximation is expected to be valid if
\begin{align}
    \frac{\hbar}{2m}\|\partial_\mu \Psi\| \ll \|\Psi\|\,.
\end{align}
The derivative of $\Psi$ can be estimated using Eq.~\eqref{eq:tau-derivative-of-Psi}.  Equation~\eqref{eq:what-should-be-0} indicates
that $\partial_\mu v_\nu$ are of the same order as $(e/2m)F_{\mu\nu}$ if they do not diverge.  
Since $(e/m)F_{\mu\nu}\sim a_0 \omega_0$, we find
\begin{align}
  \left\|\frac{d\Psi}{d\tau}\right\| \sim a_0\omega_0\|\Psi\|\,. \label{eq:estimate-for-tau}
\end{align}
We note that, since the coordinate time is $\sim a_0\tau$, the change in $\|\Psi\|$ in one cycle is of the same order as $\|\Psi\|$ 
itself beacause $a_0\tau\times\omega_0 \sim 1$ in one cycle.

Next, if $\lambda$ is a coordinate on the $\tau$-constant hypersurface, then from Eq.~\eqref{eq:tau-derivative-of-Psi} we find
\begin{align}
    \frac{d\ }{d\tau}\left(\frac{\partial\Psi}{\partial\lambda}\right)
    & = - \frac{1}{2}\left(\partial_\mu v^\mu + \frac{e}{2m}\gamma^{\mu\nu}F_{\mu\nu}\right)\frac{\partial\Psi}{\partial\lambda}\notag \\
    & \quad - \frac{\partial\ }{\partial\lambda}\left(\partial_\mu v^\mu + \frac{e}{2m}\gamma^{\mu\nu}F_{\mu\nu}\right)\Psi\,.
    \label{eq:estimate-other-than-tau}
\end{align}
If we choose $\lambda$ to be the proper length between the $\lambda$-constant surfaces, we have
\begin{align}
\frac{\partial\ }{\partial\lambda}\left(\partial_\mu v^\mu + \frac{e}{2m}\gamma^{\mu\nu}F_{\mu\nu}\right)\sim a_0\omega_0^2\,,
\end{align}
since $F_{\mu\nu}$ oscillates with angular frequency $\omega_0$.
Hence, in one cycle we have at most
\begin{align}
    \left\|\frac{\partial\Psi}{\partial\lambda}\right\| \sim a_0\omega_0 \|\Psi\|\,, \label{eq:estimate-for-lambda}
\end{align}
with possible Lorentz-boost factor of the order of $a_0$ incorporated.
[Any enhancement of $\|\partial\Psi/\partial\lambda\|$ due to the first term on the right-hand side of Eq.~\eqref{eq:estimate-other-than-tau} in one cycle is estimated to be of the same order as $\|\partial\Psi/\partial\lambda\|$ (see the comment after
Eq.~\eqref{eq:estimate-for-tau}). Also, if this term 
enhances $\|\partial\Psi/\partial\lambda\|$, then $\|\Psi\|$ should be enhanced in the same way.  
Hence we expect that it will not affect
the ratio $\|\partial\Psi/\partial\lambda\|/\|\Psi\|$ significantly.]  
Then, the derivative $\partial_\mu\Psi$ with respect to a Cartesian coordinate is estimated as
\begin{align}
    \|\partial_\mu \Psi\| \sim a_0^2\omega_0 \|\Psi\|\,,
\end{align}
by taking any Lorentz-boost factor, which is of the order of $a_0$, into account.  Thus, the condition for the validity of the WKB 
approximation is
\begin{align}
    \frac{\hbar}{m}a_0^2\omega_0 \approx \left(\frac{a_0}{700}\right)^2 \ll 1\,,
\end{align}
where we have let $\hbar\omega_0 = 1$eV and $m = 511$keV.  This is satisfied if $a_0 \approx 30$.

\section*{Appendix II: Emission probability amplitude}
\label{appII}
We expand the electron field as
\begin{align}
\begin{split}
&  \psi(t,\mathbf{x}) = \sum_{\mathbf{p},\alpha} \left[ u_{(\mathbf{p},\alpha)}(x) b_{(\mathbf{p},\alpha)} +  v_{(\mathbf{p},\alpha)}(x) d^\dagger_{(\mathbf{p},\alpha)} \right]\,,
\end{split}
\end{align}
where
\begin{align}
\begin{split}
& \int d^3 \mathbf{x} \; u^\dagger_{(\mathbf{p},\alpha)} u_{(\mathbf{p},\beta)} = \int d^3 \mathbf{x} \; v^\dagger_{(\mathbf{p},\alpha)} v_{(\mathbf{p},\beta)} = \delta_{\alpha \beta} \,, \\
       & \int d^3 \mathbf{x} \; u^\dagger_{(\mathbf{p},\alpha)} v_{(\mathbf{p},\beta)} = 0 \,,
\end{split}
\end{align}
and
\begin{align}
    \left\{ b_{(\mathbf{p},\alpha)}, b_{(\mathbf{p},\beta)}^\dagger \right\} = \left\{ d_{(\mathbf{p},\alpha)}, d_{(\mathbf{p},\beta)}^\dagger \right\} = \delta_{\alpha \beta}\,,
\end{align}
with all other anti-commutators vanishing.  Then, the equal-time anti-commutation relation,
\begin{align}
    \{\psi_a(t,\mathbf{x}),\psi^\dagger_b(t,\mathbf{x}')\} = \delta^{(3)}(\mathbf{x}-\mathbf{x'})\delta_{ab}\,,
\end{align}
is satisfied. We choose a wave-packet solution, $u_{(\mathbf{p},\alpha)}(t,\mathbf{x})$, peaked near a classical world line but approximately with a definite momentum and spin as one of these basis states.
(Thus, this wave packet behaves like a classical electron.) The index $\alpha$ specifies the spin. The WKB solutions constructed in the previous subsection satisfies $\psi^\dagger \psi = 2p_0$.  Thus, this wave-packet solution can be written
approximately as
\begin{align}
    u_{(\mathbf{p},\alpha)}(t,\mathbf{x}) = \sqrt{\frac{p_0+m}{2p_0}}\begin{pmatrix} s_\alpha \\ \frac{\mathbf{p}\cdot\boldsymbol{\sigma}}{p_0+m}\,s_\alpha \end{pmatrix}G(t,\mathbf{x})\,,
\end{align}
where $s_\alpha$ satisfies the Thomas-BMT equation, \cref{eq:Thomas-BMT}.  The function $G(t,\mathbf{x})$ is peaked about a classical world line and satisfies
\begin{align}
    \int d^3\mathbf{x} |G(t,\mathbf{x})|^2 = 1\,.
\end{align}
That is, although $G(t,\mathbf{x})$ is a smooth function, we may assume that $|G(t,\mathbf{x})|^2$ is well approximated by $\delta^{(3)}(\mathbf{x} - \mathbf{x}(t))$

The axion field is expanded as
\begin{align}
    \phi(x) & = \int \frac{d^3\mathbf{k}}{(2\pi)^32k_0}\left( a_{\mathbf{k}}e^{-ik\cdot x} + a_{\mathbf{k}}^\dagger e^{ik\cdot x}\right)\,,
\end{align}
where
\begin{align}
    \left[ a_{\mathbf{k}},a_{\mathbf{k}'}^\dagger\right] = (2\pi)^32\hbar k_0\delta^{(3)}(\mathbf{k}-\mathbf{k}')\,.
\label{eq:comm-rel-axion-supp}    
\end{align}
Although the axion mass $m_a$ is nonzero, we let $m_a = 0$, assuming that it is
very small.
The term describing the interaction between the electron
field and the axion field, $\phi$, in the Lagrangian density is
\begin{align}
    \mathcal{L}_{\textrm{int}}
    & = -\frac{\hbar g_{ae}}{2m}\partial_\mu \phi \overline{\psi}\gamma_5 \gamma^\mu \psi\,.
\end{align}
The interaction Hamiltonian density can be found following the standard procedure as
\begin{align}
    \mathcal{H}_{\textrm{int}} = 
     \frac{\hbar g_{ae}}{2m}\partial_\mu \phi \overline{\psi}\gamma_5 \gamma^\mu \psi
     + \frac{1}{2}\left(\frac{\hbar g_{ae}}{2m}\right)^2
     (\overline{\psi}\gamma_5\gamma^0\psi)^2\,.
\end{align}
To lowest order in perturbation theory we may neglect the four-fermi interaction term.  For this reason
we adopt the first term as the interaction Hamiltonian density from now on. 

Let the annihilation operator corresponding to the wave-packet solution $u_{(\mathbf{p},\alpha)}(t,\mathbf{x})$ be denoted by $b_{\alpha}$. 
(We omit ``$\mathbf{p}$''
to simplify the notation.)  The initial and final states are
$b_{\alpha}^\dagger\ket{0}$, and $a_{\mathbf{k}}^\dagger b_{\beta}^\dagger\ket{0}$, where $a_{\mathbf{k}}^\dagger$ is the creation operator for the
axion with momentum $\mathbf{k}$.  We assume that the solutions corresponding to $b_{\alpha}^\dagger\ket{0}$ and $b_{\beta}^\dagger\ket{0}$ are approximately 
the same except possibly for
the spin states.  That is, $s_\alpha$ and $s_\beta$ may or may not be the same, but they have approximately the same spatial wave function. If the initial state is
$b_\alpha^\dagger\ket{0}$, then the final one-axion-emission state is, with
$:\cdots:$ denoting normal-ordering,
\begin{widetext}
\begin{align}
\begin{split}
\ket{f} &=  - \frac{i}{\hbar}\int d^4 x\,:\!\mathcal{H}_{\mathrm{int}}(x)\!:b_{\alpha}^\dagger\ket{0}\\ & =  \frac{g_{ae}}{2m}\int \frac{d^3\mathbf{k}}{(2\pi)^32k_0} \sum_{\beta}\int d^4 x\, \overline{u_{\beta}}\gamma_5\slashed{k}u_{\alpha}e^{ik\cdot x}
    a_{\mathbf{k}}^\dagger b_{\beta}^\dagger\ket{0} \\
    & = \frac{g_{ae}}{2m} \int \frac{d^3\mathbf{k}}{(2\pi)^32k_0} \sum_\beta\int \frac{dt}{2p_0}
     \overline{\Phi_\beta}\gamma_5\slashed{k}\Phi_\alpha\,e^{ik_0t - i\mathbf{k}\cdot\mathbf{x}(t)} a_{\mathbf{k}}^\dagger b_{\beta}^\dagger\ket{0} \\
     & = \frac{g_{ae}}{4m^2} \int \frac{d^3\mathbf{k}}{(2\pi)^3 2k_0} \sum_\beta \int d\tau
     \overline{\Phi_\beta}\gamma_5\slashed{k}\Phi_\alpha\,e^{ik\cdot x(\tau)} a_{\mathbf{k}}^\dagger b_{\beta}^\dagger\ket{0}\,,
\end{split}
\end{align}
where
\begin{align}
\Phi_\alpha 
= \sqrt{p_0+m}\begin{pmatrix} s_\alpha \\ \frac{\mathbf{p}\cdot\boldsymbol{\sigma}}{p_0+m}\,s_\alpha\end{pmatrix}\,.
\label{eq:sol-for-Phi}
\end{align}
\end{widetext}
We find
\begin{align}
\overline{\Phi_{\beta}}\gamma_5\slashed{k}\Phi_{\alpha}
& = 2s_\beta^\dagger\left( m \mathbf{k}\cdot\boldsymbol{\sigma} - \frac{k\cdot p + mk}{p_0+m}\mathbf{p}\cdot\boldsymbol{\sigma}  \right)s_\alpha\,,
\end{align}
Since $\bra{0}b_{\beta}b_{\beta}^\dagger\ket{0} = 1$, 
using \cref{eq:comm-rel-axion-supp} we find
the probability for emission to be
\begin{align}
    P_{\mathrm{em}} = \hbar \int\frac{d^3\mathbf{k}}{(2\pi)^32k_0}
    \sum_{\beta}\left|\mathcal{A}_{(\mathbf{p},\mathbf{k},\beta,\alpha)}\right|^2\,,
\label{eq:prob-supp}    
\end{align}
\begin{widetext}
where
\begin{align}
 \   \mathcal{A}_{(\mathbf{p},\mathbf{k},\beta,\alpha)} & = \frac{g_{ae}}{2m}\int d\tau \,s_\beta^\dagger\left[ \mathbf{k}\cdot\boldsymbol{\sigma} - \frac{k\cdot p + mk}{m(p_0+m)}\mathbf{p}\cdot\boldsymbol{\sigma}  \right]s_\alpha\,e^{ik\cdot x(\tau)}\notag \\
    & = \frac{i g_{ae}}{2m}\int d\tau\frac{d\ }{d\tau}\left\{ (k\cdot v)^{-1} s_\beta^\dagger\left[ \mathbf{k}\cdot\boldsymbol{\sigma} - \frac{k\cdot p + mk}{m(p_0+m)}\mathbf{p}\cdot\boldsymbol{\sigma}  \right]s_\alpha\right\}\,e^{ik\cdot x(\tau)} \notag \\
    & = \frac{i g_{ae}}{2m}\int d\tau\, \frac{e^{ik\cdot x(\tau)}}{(k\cdot v)^2}
    s_\beta^\dagger(\tau)\mathbf{Q}(\tau)\cdot\boldsymbol{\sigma}s_\alpha(\tau)\,,
\label{eq:amplitude-supp}    
\end{align}
\end{widetext}
where in the second step we integrated by parts and  defined
\begin{align}
     \mathbf{Q} & = \mathbf{V}
    - (k\cdot a)\mathbf{k}
    - \left[V_0 - (k\cdot a)k_0\right]\frac{\mathbf{p}}{p_0+m}\,,
    \label{eq:simplified-Q}
\end{align}
with
\begin{align}
    V^\mu = \frac{e}{m}(k\cdot v)F^{\mu\nu}k_\nu\,,
\end{align}
which can be re-expressed using \cref{eq:F-in-terms-of-EB} as
\begin{align}
    V_0 & = \frac{e}{m}(k\cdot v)\mathbf{k}\cdot\mathbf{E}\,,\\
    \mathbf{V} & = \frac{e}{m}(k\cdot v)
    (k_0\mathbf{E} + \mathbf{k}\times\mathbf{B})\,.
\end{align}
Here, $a^\mu = dv^\mu/d\tau$ is the proper acceleration.
A useful formula for finding $\mathbf{Q}$ is
\begin{align}
    \frac{d\ }{d\tau}\left(\frac{\mathbf{p}}{p_0+m}\right)
    =\frac{2\mathbf{F}\times\mathbf{p} - e\mathbf{E}}{p_0+m}\,,
    \label{eq:useful-formula}
\end{align}
which can be derived from the Lorentz-force equations~\eqref{eq:Lorentz-force-EB-1}.
It is also useful to note that the Lorentz-force equations can be given as
\begin{align}
    a^\mu & = - \frac{e}{m}F^{\mu\nu}v_\nu\,,
    \label{eq:formula-for-acceleration}
\end{align}
or
\begin{align}
    a_0 & = - \frac{e}{m}\mathbf{v}\cdot \mathbf{E}\,,
    \label{eq:acceleration-time}\\
    \mathbf{a} & = - \frac{e}{m}(v_0 \mathbf{E} + \mathbf{v}\times\mathbf{B})\,.\label{eq:acceleration-space}
\end{align}


\subsection{One-dimensional motion}

Suppose that the electric field $\mathbf{E}$ is parallel to the $z$-direction and $\mathbf{B}=0$.  Then, the motion can occur along the
$z$-axis.  From \cref{eq:Thomas-BMT} we find that the spin is time independent.  
Let the initial spin $s_\alpha$ be pointing in the $z$-direction.  If the final spin $s_\beta$ is pointing in the
same direction, then the corresponding amplitude, i.e., the spin-non-flip amplitude,
is
\begin{align}
    \mathcal{A}_{(p_z,\mathbf{k})}^{\textrm{nf}}
    & = \frac{i g_{ae}}{2m}\int d\tau\,\frac{e^{ik\cdot x}}{(k\cdot v)^2} Q_z(\tau)\,,
\end{align}
where~\footnote{If we do not make the approximation $m_a \approx 0$, where $m_a$ is
the axion mass, then $k_\perp^2$
here is replaced by $k_\perp^2 + m_a^2/\hbar^2$.}
\begin{align}
    Q_z(\tau) & = - k_\perp^2 a(\tau)\,,
\end{align}
with the acceleration $a(\tau)$ defined by
\begin{align}
    a(\tau) = - \frac{e}{m}E_z(\tau)\,.
\end{align}
Hence, the spin non-flip axion emission probability is
\begin{align}
    P_{\mathrm{em}}^{\mathrm{nf}} = \frac{\hbar g_{ae}^2}{4m^2}\int \frac{d^3\mathbf{k}}{(2\pi)^32k_0} k_\perp^4
    \left|\int d\tau\, \frac{a}{(k\cdot v)^2}e^{ik\cdot x}\right|^2\,.
\end{align}

If the final spin is in the opposite direction, then
    \begin{align}
    \mathcal{A}_{(p_z,\mathbf{k})}^{\textrm{f}}
    & = \frac{i g_{ae}}{2m}\int d\tau\,\frac{e^{ik\cdot x}}{(k\cdot v)^2} [Q_x(\tau) + iQ_y(\tau)]\,,
\end{align}
where
\begin{align}
    Q_x(\tau) + iQ_y(\tau) & = - (k\cdot a)(k_x + ik_y)\,.
\end{align}
Hence the spin-flip emission probability is
\begin{align}
    P_{\textrm{em}}^{\textrm{f}}
    = \frac{\hbar g_{ae}^2}{4m^2}\int \frac{d^3\mathbf{k}}{(2\pi)^32k_0}\,k_\perp^2\left| 
    \int d\tau\,\frac{k\cdot a}{(k\cdot v)^2} e^{ik\cdot x}\right|^2\,.
\end{align}

\subsection{Two-dimensional motion}

If the electric field is on the $xz$-plane and the
magnetic field is in the $y$-direction, then the
electron motion can occur on the $xz$-plane. In this
case it is convenient to choose the initial spin in the
$y$-direction since \cref{eq:Thomas-BMT} reads
\begin{align}
    \frac{ds_{\pm}}{d\tau} = \mp \frac{ie}{2m}\left(B_y - \frac{v_zE_x- v_x E_z}{v_0 + 1}\right) s_\pm\,,
\end{align}
where $\sigma_y s_{\pm} = \pm s_{\pm}$, so that
\begin{align}
\begin{split}
&s_{\pm}(\tau)  \\
& =\exp\left[ \mp\frac{ie}{2m}\int_{\tau_i}^\tau \left(B_y - \frac{v_zE_x- v_xE_z}{v_0+1}\right)d\xi\right]s_{\pm}(\tau_i)\,.
\end{split}
\end{align}
Then, the spin-non-flip emission amplitude is 
\begin{align}
    P_{\mathrm{em}}^{\mathrm{nf}} =\frac{\hbar g_{ae}^2}{4m^2}
    \int \frac{d^3\mathbf{k}}{(2\pi)^32k_0}k_y^2
    \left|\int d\tau\,e^{ik\cdot x}\frac{k\cdot a}{(k\cdot v)^2}\right|^2\,.
\end{align}
The spin-flip amplitude is
\begin{align}
    P_{\mathrm{em}}^{\mathrm{f}} & = \frac{\hbar g_{ae}^2}{4m^2}
    \int \frac{d^3\mathbf{k}}{(2\pi)^32k_0}
    \left| \int d\tau\, e^{ik\cdot x} e^{\mp if(\tau)}\frac{Q_z \pm i Q_x}{(k\cdot v)^2} \right|^2
\end{align}
where
\begin{align}
    Q_z \pm i Q_x & = -\frac{e}{m}(k\cdot v)[ k_0(E_z\pm i E_x) + (k_x\mp ik_z)B_y]\notag \\
    & \quad \quad - (k\cdot a)(k_z\pm ik_x)\notag \\
    & \quad - \left[ \frac{e}{m}(k\cdot v)\mathbf{k}\cdot\mathbf{E} - (k\cdot a)k_0\right]\frac{v_z\pm iv_x}{v_0+1}\,,\\
    f(\tau) & = \frac{e}{m}\int_{\tau_0}^\tau \left[ B_y(\xi) - \frac{(\mathbf{v}(\xi)\times\mathbf{E}(\xi))_y }{v_0(\xi)+1} \right]d\xi\,. 
\end{align}
\section*{Appendix III: The Feynman-diagram derivation of the emission probability in the weak-field limit}
\label{appIII}
Here we investigate the emission process in the weak-field limit but without
assuming that the energy of the axion emitted is of lower order in $\hbar$ compared
to the energy of the electron.  We let $\hbar=1$.

\begin{figure}[ht]
\centering
\begin{tikzpicture}
\fill (-3,0) circle (1.5mm);
\draw[very thick] (-2,-2) node[right] {$e^{-}$} -- (-3,0) node[left] {$A_\mu \; $} -- (-2,2) node[left] {$e^{-}$};
\fill (2,0) circle (1.5mm);
\draw[very thick] (3,-2) node[right] {$e^{-}$} -- (2,0) node[left] {$A_\mu \; $} -- (3,2)node[right] {$e^{-}$};
\draw[very thick] (0,0) node[cross=1.5mm,rotate=45]{};
\draw[very thick,dashed] (-2.5,1) -- (-1.5,1.5) node[right] {$a$};
\draw[very thick,dashed] (2.5,-1) -- (3,0)node[right] {$a$};
\end{tikzpicture}
\caption{The Feynman diagrams for the axion emission: The solid line represents the electron and the dashed line
represents the axion.  
The dot represents the interaction of the electron with the external potential.}
\label{fig:feynman-diagrams}
\end{figure}
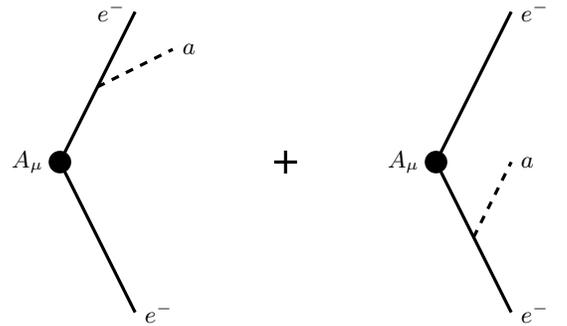

Assume that the external electromagnetic potential $A_\mu$ is nonzero only in a finite
spacetime region and smooth.  Then, the spacetime Fourier transform,
\begin{align}
    \widetilde{A}_\mu(q) & = \int e^{iq\cdot x}A_\mu(x)\,d^4 x\,,
\end{align}
exists.
The electron propagator is
\begin{align}
    D(p) & = \frac{i(\slashed{p} + m)}{p^2 - m^2+ i\epsilon}\,.
\end{align}
The relevant Feynman diagram is shown in Fig.~\ref{fig:feynman-diagrams}.
With the initial spinor $u^{(i)}(p_i)e^{-ip_i\cdot x}$ and the
final spinor $u^{(f)}(p_f)e^{-ip_f\cdot x}$, the amplitude for the emission of the axion
with $4$-momentum $k^\mu$ is
\begin{widetext}
\begin{align}
    \mathcal{M} & = \frac{eg_{ae}}{2m}\overline{u^{(f)}(p_f)}
    [\gamma_5\slashed{k}D(p_f+k)\gamma^\mu - \gamma^\mu D(p_i-k)\gamma_5\slashed{k}]
    u^{(i)}(p_i)\widetilde{A}_\mu(q)\,,
\end{align}
\end{widetext}
where $u^{(i)}(p)$ and $u^{(f)}(p)$ are normalized so that
\begin{align}
    \sum_{\textrm{spin}}u^{(i)}(p)\overline{u^{(i)}(p)} = \slashed{p} + m\,,
\end{align}
and similarly for $u^{(f)}(p)$.  We have defined
    $q= p_f + k - p_i$.
Formally, the initial state, $b_{(\mathbf{p}_i,\alpha)}^\dagger\ket{0}$, is normalized as 
$\bra{0}b_{(\mathbf{p}_i,\alpha)}b_{(\mathbf{p}_i,\alpha)}^\dagger\ket{0} = 
(2\pi)^3 2p_i^0\delta^{(3)}(\mathbf{0}) = 2p_i^0 V$,
where $V$ is the (infinite) volume of the space. The flux is
$2p_i^0 [\|\mathbf{p}_i\|/p_i^0] = 2\|\mathbf{p}_i\|$.  Define
\begin{align}
    \Sigma= \frac{1}{2p_i^0}
    \int \frac{d^3\mathbf{k}}{(2\pi)^32k_0}
    \frac{d^3\mathbf{p}_f}{(2\pi)^32p_f^0}|\mathcal{M}|^2\,,
\end{align}
which we call the \textit{interaction volume}.
If the electron beam has the number density $\rho$, then the emission probability is given by $\rho\Sigma$.

By averaging over the initial spin and summing over the final spin, we find
\begin{align}
    \frac{1}{2}\sum_{\textrm{spin\ sum}}|\mathcal{M}|^2
    & =e^2 \mathcal{M}^{\mu\nu}\widetilde{A}^*_\mu(q)\widetilde{A}_\nu(q)\,,
\end{align}
where
\begin{widetext}
\begin{align}
    \mathcal{M}^{\mu\nu}
    & = -\frac{1}{2}\mathrm{Tr}
    \left\{ (\slashed{p}_f+m)
    \left[ \frac{\gamma_5\slashed{k}(\slashed{p}_f+m)\gamma^\mu}{2p_f\cdot k} - \frac{\gamma^\mu(\slashed{p}_i+m)\gamma_5\slashed{k}}{2p_i\cdot k}\right]
    (\slashed{p}_i+m)\left[
    \frac{\gamma^\nu(\slashed{p}_f+m)\slashed{k}\gamma_5}{2p_f\cdot k}
    - \frac{\slashed{k}\gamma_5(\slashed{p}_i+m)\gamma^\nu}{2p_i\cdot k}
    \right]\right\}\,.
\end{align}
\end{widetext}
The overall minus sign is due to the insertion of $\gamma_5$ and $\gamma^0\gamma_5\gamma^0 = -\gamma^5$.
After a tedious but straightforward calculation we find
\begin{align}
    \frac{1}{2}\sum_{\textrm{spin\ sum}}|\mathcal{M}|^2
    = \frac{4e^2}{(k\cdot v_f)(k\cdot v_i)}k^\mu\widetilde{F}_{\mu\nu}(q)\widetilde{F}^{\nu\lambda}(q)k_\lambda\,,
\end{align}
where
\begin{align}
    \widetilde{F}_{\mu\nu}(q) & = \int d^4 x\, e^{iq\cdot x}[\partial_\mu A_\nu(x) - \partial_\nu A_{\mu}(x)] \notag \\
    & = -i[q_\mu \widetilde{A}_\nu (q) -  q_\nu \widetilde{A}_\mu (q)]\,.
\end{align}
Thus, we find
\begin{align}
    \Sigma = \frac{e^2g_{ae}^2}{2p_i^0 m^2}
        \int \frac{d^3\mathbf{k}}{(2\pi)^32k_0}
    \frac{d^3\mathbf{p}_f}{(2\pi)^32p_f^0}
    \frac{k^\mu \widetilde{F}^*_{\mu\nu}(q)\widetilde{F}^{\nu\lambda}(q)k_\lambda}{(k\cdot v_f)(k\cdot v_i)}\,.  \label{eq:sigma-tilde}
\end{align}

Now, we consider the case where the electromagnetic field is only $t$-dependent,
i.e., space-independent with the assumption that the characteristic frequency of the electromagnetic field
is much smaller than the Compton wavelength of the electron so that, typically, $k_0 \ll \|\mathbf{p}_i\|, \|\mathbf{p}_f\|$.  The aim is to show
that the Feynman-diagram method gives the same
result as our WKB result in this limit.

The Fourier transform of the electromagnetic field becomes
\begin{align}
    \widetilde{F}_{\mu\nu}(q) & = (2\pi)^3\delta^{(3)}(\mathbf{p}_f + \mathbf{k} - \mathbf{p}_i)\hat{F}(p_f^0 - k_0 + p_i^0)\,,
    \label{eq:F-for-space-indep}
\end{align}
where
\begin{align}
    \hat{F}(p_f^0+k_0 - p_i^0) = \int e^{i(p_f^0 + k - p_i^0)t}F_{\mu\nu}(t)\,dt\,.
\end{align}
Then, by using the momentum conservation, we have
\begin{align}
    p_f^0 - p_i^0 = - \mathbf{k}\cdot\boldsymbol{\beta},
\end{align}
to first order in $k_0$, where $\boldsymbol{\beta}= \mathbf{p}_i/p_i^0$.  Thus, we find
\begin{align}
\hat{F}(p_f^0+k_0 - p_i^0) = v_0\int e^{ik\cdot x(\tau)}F_{\mu\nu}(\tau)\,d\tau\,,
\end{align}
where $v_0 = v_i^0 = v_f^0$ at lowest order in
$\hbar$.  Substituting \cref{eq:F-for-space-indep} into
\cref{eq:sigma-tilde} and using the formal equality $(2\pi)^3\delta^{(3)}(\mathbf{0}) = V$
we find
\begin{align}
    \Sigma = V\frac{e^2g_{ae}^2}{4m^4}
        \int \frac{d^3\mathbf{k}}{(2\pi)^32k_0}
    \frac{k^\mu \widetilde{F}^*_{\mu\nu}(q)\widetilde{F}^{\nu\lambda}(q)k_\lambda}{(k\cdot v)^2}\,,
\end{align}
where $v^\mu = v_f^\mu = v_i^\mu$ at this order.
Then, since the number density of one electron in volume $V$ is $\rho=1/V$, the probability of axion emission is
\begin{align}
    P_{\mathrm{em}}^{(\mathrm{av})} & = \Sigma/V \notag \\
    & = \frac{\hbar e^2g_{ae}^2}{4m^4}\int \frac{d^3\mathbf{k}}{(2\pi)^32k_0} \notag \\
 &  \times   \int d\tau'\int d\tau \frac{e^{-ik\cdot [x(\tau')-x(\tau)]}}{(k\cdot v)^2}k_\mu F^{\mu\nu}(\tau')F_{\nu\lambda}(\tau)k^\lambda\,. \label{eq:weak-field-approx}
\end{align}

Now we show that \cref{eq:weak-field-approx}
agrees with the weak-field limit of 
\cref{eq:prob-supp}.
Note that the spin may be taken to be constant
in the weak-field limit.  Then
\begin{align}
\begin{split}
& \frac{1}{2}\sum_{\alpha,\beta} s_\beta^\dagger \mathbf{Q}(\tau')\cdot\boldsymbol{\sigma}s_\alpha s_\alpha^\dagger\mathbf{Q}(\tau)\cdot \boldsymbol{\sigma}s_\beta \\
& = \frac{1}{2}\textrm{Tr}[\mathbf{Q}(\tau')\cdot\boldsymbol{\sigma}\mathbf{Q}(\tau)\cdot
\boldsymbol{\sigma}] = \mathbf{Q}(\tau')\cdot\mathbf{Q}(\tau)\,.
\end{split}
\end{align}
In the expression~\eqref{eq:simplified-Q} for $\mathbf{Q}$ we may approximate the $4$-momentum $p^\mu = mv^\mu$ to
be constant and treat only the electromagnetic field to be time-dependent.
Since 
\begin{align} 
[V^\mu - (k\cdot a)k^\mu]v_\mu = 0\,,
\end{align}
i.e.,
\begin{align}
    [\mathbf{V}(\tau) - (k\cdot a(\tau))\mathbf{k}]\cdot \mathbf{v} = [V_0(\tau) - (k\cdot a(\tau))k_0]v_0\,,
\end{align}
and $\mathbf{p}^2 = p_0^2 - m^2$, we find
\begin{align}
    \mathbf{Q}(\tau')\cdot\mathbf{Q}(\tau) & =  -V^\mu(\tau')V_\mu(\tau) \notag \\
    & = \frac{e^2}{m^2}(k\cdot v)^2 k_\mu F^{\mu\nu}(\tau')F_{\nu\lambda}(\tau)k^\lambda\,,  \label{eq:Q-squared-weak-field}
\end{align}
in the weak-field limit,
where we also used $V^\mu(\tau) k_\mu = 0$. Then, we find
that the weak-field approximation of \cref{eq:prob-supp} 
indeed agrees with \cref{eq:weak-field-approx}.

\section*{Appendix IV: Energy of emitted axions}
\label{appIV}
Here we derive the formula for the energy of the emitted axions.
As is the case with the Larmor radiation (see,
e.g., Ref.~\cite{Higuchi:2009ms}), the energy emitted takes a local form
in the sense that it is a single integral, rather
than a double integral, over the proper time.  That is, there
are no interference terms between different proper times.  This is in contrast
with the number of emitted axions, which has interference terms of this kind.

Define
\begin{align}
\begin{split}
&\widetilde{\mathbf{Q}}  = k_0^{-2}\mathbf{Q}\,, \quad 
    n^\mu  = \frac{k^\mu}{k_0}\,, \\
   & \widetilde{V}^\mu 
     = F^{\mu\nu}n_\nu\,.
\end{split}     
\end{align}
Thus,
\begin{align}
\begin{split}
\widetilde{\mathbf{Q}}
    & = \frac{e}{m}(n\cdot v)\widetilde{\mathbf{V}}
    - (n\cdot a)\mathbf{n} \\
    &+ \frac{\mathbf{p}}{p_0+m}\left[ n\cdot a - \frac{e}{m}
    (n\cdot v) \widetilde{V}_0\right]\,.
\end{split}
\end{align}
Then, the energy emitted is obtained from \cref{eq:prob-supp} as
\begin{widetext}
\begin{align}
    E_{\beta\alpha}
    & = \frac{g_{ae}^2\hbar}{16\pi^2 m^2}
    \int \frac{d\Omega}{4\pi} 
    \int_0^\infty dk\,k^2
    \int \frac{d\tau}{[n\cdot v(\tau)]^2}\int \frac{d\tau'}{[n\cdot v(\tau')]^2} 
    s_\beta^\dagger(\tau)\widetilde{\mathbf{Q}}(\tau)\cdot\boldsymbol{\sigma}
    s_\alpha(\tau)s_\alpha^\dagger(\tau')
    \widetilde{\mathbf{Q}}(\tau')\cdot\boldsymbol{\sigma}s_\beta(\tau')
    e^{ik\cdot [x(\tau)-x(\tau')]}\,.
\end{align}
We define $\xi = t - \mathbf{n}\cdot \mathbf{x}$~\cite{Higuchi:2005an}. Then
$d\xi/d\tau = n\cdot v$ so that $d\tau/(n\cdot v)^2 = d\xi/(n\cdot v)^3$.
Thus
\begin{align}
E_{\beta\alpha}
    & = \frac{g_{ae}^2\hbar}{16\pi^2 m^2}
    \int \frac{d\Omega}{4\pi} 
    \int_0^\infty dk
    \int d\xi\int d\xi'
   \notag \\
   & \quad \times\frac{d\ }{d\xi'}\left\{\frac{1}{[n\cdot v(\tau)]^3} s_\beta^\dagger(\tau)\widetilde{\mathbf{Q}}(\tau)\cdot\boldsymbol{\sigma}
        s_\alpha(\tau)\right\}
\frac{d\ }{d\xi'}\left\{\frac{1}{[n\cdot v(\tau')]^3}s_\alpha^\dagger(\tau')
    \widetilde{\mathbf{Q}}(\tau')\cdot\boldsymbol{\sigma}s_\beta(\tau')\right\}
    e^{ik(\xi-\xi')}\,,
\end{align}
where we have integrated by parts.
\end{widetext}
If we sum over the final spin states and average over the initial spin states, then the integrand is symmetric under the exchange $\xi\leftrightarrow \xi'$.
This symmetry can be used to extend
the range of the $k$-integral to $(-\infty,\infty)$.  Thus, we find the total energy emitted with the initial spin averaged over as
\begin{align}
\langle E \rangle
    & = \frac{g_{ae}^2\hbar}{32\pi m^2}
    \int \frac{d\Omega}{4\pi}
    \int \frac{d\tau}{n\cdot v}\notag \\
    & \quad \times \sum_{\alpha,\beta}
   \left| \frac{d\ }{d\tau}\left[\frac{1}{[n\cdot v(\tau)]^3} s_\beta^\dagger(\tau)\widetilde{\mathbf{Q}}(\tau)\cdot\boldsymbol{\sigma}
    s_\alpha(\tau)\right]\right|^2\,.
\end{align}

Now,
\begin{align}
\frac{d\ }{d\tau}\left[\frac{1}{[n\cdot v(\tau)]^3} s_\beta^\dagger(\tau)\widetilde{\mathbf{Q}}(\tau)\cdot\boldsymbol{\sigma}
    s_\alpha(\tau)\right]
    & = s_\beta^\dagger \mathbf{S}\cdot\boldsymbol{\sigma}s_\alpha\,,
\end{align}
where
\begin{align}
    \mathbf{S} & = 
    - \frac{3n\cdot a}{(n\cdot v)^4}\widetilde{\mathbf{Q}}
    + \frac{1}{(n\cdot v)^3}
    [\dot{\widetilde{\mathbf{Q}}}
    - 2\mathbf{F}\times\widetilde{\mathbf{Q}}]\,,
\end{align}
where the dot denotes the $\tau$-derivative and $\mathbf{F}$ is defined in \cref{eq:definition-of-F}.
Then,
since $\sum_\alpha s_\alpha(\tau)s_\alpha^\dagger(\tau) = \mathbbm{1}$ 
for all $\tau$, we find
\begin{align}
    \langle E\rangle 
    & = 
     \frac{\hbar^2g_{ae}^2}{32\pi m^2}
    \int \frac{d\Omega}{4\pi}
    \int \frac{d\tau}{n\cdot v}
    \mathrm{Tr}\left[ (\mathbf{S}\cdot\boldsymbol{\sigma})^2\right]\notag \\
    & =  \frac{g_{ae}^2\hbar}{16\pi m^2}
    \int \frac{d\Omega}{4\pi}
    \int \frac{d\tau}{(n\cdot v)^7} \notag \\
    &  \times
    \left[ \frac{9(n\cdot a)^2}{(n\cdot v)^2}\widetilde{\mathbf{Q}}^2 
    - \frac{3n\cdot a}{n\cdot v}\frac{d\ }{d\tau}\widetilde{\mathbf{Q}}^2
    +\|\dot{\widetilde{\mathbf{Q}}} - 2\mathbf{F}\times\widetilde{\mathbf{Q}}\|^2\right]\,. \label{eq:still-complicated}
\end{align}
To simplify the integrand in this equation, it is useful to define
\begin{align}
    U^\mu & = (n\cdot a)n^\mu  -\frac{e}{m}(n\cdot v)\widetilde{V}^\mu \notag \\
    & = - \frac{e}{m}\left( n_\alpha F^{\alpha\beta}v_\beta\, n^\mu 
    - n_\alpha F^{\alpha\mu} n^\nu v_\nu \right)\,. \label{eq:U-expression}
\end{align}
Then, 
\begin{align}
\widetilde{\mathbf{Q}} = - \mathbf{U} + \frac{m\mathbf{v}}{p_0+m}U_0\,.
\end{align}
Equation~\eqref{eq:Q-squared-weak-field} is translated to
\begin{align}
    \widetilde{\mathbf{Q}}^2
    & = - U^\mu U_\mu\,.
\end{align}

Next we examine $\|\dot{\widetilde{\mathbf{Q}}} - 2\mathbf{F}\times\widetilde{\mathbf{Q}}\|^2$.
We find
\begin{align}
\dot{\widetilde{\mathbf{Q}}} - 2\mathbf{F}\times\widetilde{\mathbf{Q}} 
    & = - \frac{d\mathbf{U}}{d\tau} + \frac{\mathbf{p}}{p_0+m}\frac{dU_0}{d\tau}
    + \frac{d\ }{d\tau}\left( \frac{\mathbf{p}}{p_0+m}\right)U_0
\notag \\
& \quad  + 2\mathbf{F}\times\mathbf{U} - \frac{2\mathbf{F}\times\mathbf{p}}{p_0+m}U_0\,. 
\end{align}
By \cref{eq:useful-formula} and the definition~\eqref{eq:definition-of-F}, we have
\begin{align}
    & \dot{\widetilde{\mathbf{Q}}} - 2\mathbf{F}\times\widetilde{\mathbf{Q}}\notag \\
    & = - \frac{d\mathbf{U}}{d\tau} + \frac{\mathbf{p}}{p_0+m}\frac{dU_0}{d\tau}
    - \frac{eU_0\mathbf{E}}{p_0+m} \notag \\
    & \quad 
    - \frac{e}{m}\mathbf{U}\times\mathbf{B} - \frac{e}{m(p_0+m)}
    \left[ (\mathbf{p}\cdot\mathbf{U})\mathbf{E} - (\mathbf{E}\cdot\mathbf{U})\mathbf{p}\right]\,.
\end{align}
Using $U^\mu v_\mu = 0$, i.e., $\mathbf{p}\cdot\mathbf{U} = p_0 U_0$, we find 
\begin{align}
    \dot{\widetilde{\mathbf{Q}}} - 2\mathbf{F}\times\widetilde{\mathbf{Q}}
    & = - \left[\frac{d\mathbf{U}}{d\tau} + 
    \frac{e}{m}(U_0\mathbf{E} + \mathbf{U}\times\mathbf{B})\right] \notag \\
    & \quad 
    + \frac{\mathbf{p}}{p_0+m} 
    \left[ \frac{dU_0}{d\tau} + \frac{e}{m}\mathbf{U}\cdot\mathbf{E} \right]\,.
\end{align}
Define
\begin{align}
    H^\mu & = \frac{dU^\mu}{d\tau} + \frac{e}{m}F^{\mu\nu}U_\nu\,.
\end{align}
Then,
\begin{align}
   \dot{\widetilde{\mathbf{Q}}} - 2\mathbf{F}\times\widetilde{\mathbf{Q}}
   & = - \mathbf{H} + \frac{\mathbf{p}}{p_0+m}H_0\,.
\end{align}
The equation $dv^\mu/d\tau = - (e/m)F^{\mu\nu}v_\nu$ can be used to show that
$H^\mu v_\mu = H_0v_0 - \mathbf{H}\cdot\mathbf{v} = 0$.
Then, the same calculation as that demonstrated $\widetilde{\mathbf{Q}}^2 = -U^\mu U_\mu$ shows that
\begin{align}
    \|\dot{\widetilde{\mathbf{Q}}} - 2\mathbf{F}\times\widetilde{\mathbf{Q}}\|^2
    = - H^\mu H_\mu\,.
\end{align}
Thus, we find that
\begin{align}
    \langle E\rangle & =\frac{ \hbar^2g_{ae}^2}{16\pi m^2}
    \int \frac{d\Omega}{4\pi}
    \int \frac{d\tau}{(n\cdot v)^7} \notag \\
&  \times    \left[ -\frac{9(n\cdot a)^2}{(n\cdot v)^2}U_\mu U^\mu  
    + \frac{3n\cdot a}{n\cdot v}\frac{d\ }{d\tau}(U_\mu U^\mu)
    - H_\mu H^\mu\right]\,, \label{eq:starting-point}
    \end{align}
where $U^\mu$ is given by \cref{eq:U-expression}, and
\begin{equation}
\begin{aligned}
 H^\mu 
 & = \frac{d(n\cdot a)}{d\tau}n^\mu - \frac{e}{m}(n\cdot v)\dot{F}^{\mu\nu}n_\nu\notag \\
  & \quad  - \frac{e^2}{m^2}(n\cdot v)F^{\mu\nu}F_{\nu\lambda}n^\lambda\,.
\end{aligned}
\end{equation}

When evaluating $U_\mu U^\mu$,  
we can use $n^\mu n_\mu = 0$ and $n^\mu \widetilde{V}_\mu = 0$.  Thus,
\begin{align}
    U_\mu U^\mu = -\frac{e^2}{m^2}(n\cdot v)^2 n_\mu F^{\mu\nu}F_{\nu\lambda}n^\lambda\,. 
    \end{align}
Similarly, one can simplify $H_\mu H^\mu$ as
\begin{widetext}
\begin{align}
    H_\mu H^\mu & = - \frac{2e^2}{m^2}(n\cdot v)\frac{d(n\cdot a)}{d\tau}
    n_\mu F^{\mu\nu}F_{\nu\lambda}n^\lambda 
    + \frac{e^2}{m^2}(n\cdot v)^2\left(\dot{F}^{\mu\nu} + \frac{e}{m}F^{\mu\alpha}{F_\alpha}^\nu\right)n_\nu
    \left(\dot{F}_{\mu\lambda} + \frac{e}{m}F_{\mu\alpha}{F^\alpha}_\lambda\right)n^\lambda\,.
\end{align}
Hence, the expression inside the brackets of \cref{eq:starting-point} divided by $(n\cdot v)^7$ is
\begin{align}
& \frac{1}{(n\cdot v)^7}\left[ -\frac{9(n\cdot a)^2}{(n\cdot v)^2}U_\mu U^\mu  + \frac{3n\cdot a}{n\cdot v}\frac{d\ }{d\tau}(U_\mu U^\mu)
    - H_\mu H^\mu\right] \notag \\
    &= \frac{3e^2}{m^2}\frac{(n\cdot a)^2}{(n\cdot v)^7}n_\mu F^{\mu\nu}F_{\nu\lambda}n^\lambda
- \frac{6e^2}{m^2}\frac{n\cdot a}{(n\cdot v)^6}n_\mu \dot{F}^{\mu\nu}F_{\nu\lambda}n^\lambda  + \frac{2e^2}{m^2}\frac{n\cdot \dot{a}}{(n\cdot v)^6}
    n_\mu F^{\mu\nu}F_{\nu\lambda}n^\lambda \notag \\
& \quad    - \frac{e^2}{m^2}\frac{1}{(n\cdot v)^5}\left(\dot{F}^{\mu\nu} + \frac{e}{m}F^{\mu\alpha}{F_\alpha}^\nu\right)n_\nu
    \left(\dot{F}_{\mu\lambda} + \frac{e}{m}F_{\mu\beta}{F^\beta}_\lambda\right)n^\lambda\,.
\label{eq:the-Y}
\end{align}

\subsection{Angular integral for the energy emitted}

Next we carry out the angular integral in \cref{eq:starting-point}.
All integrals can be found from
\begin{align}
\frac{1}{4\pi}\int d\Omega\, \frac{n_{\mu_1}n_{\mu_2}}{(n\cdot v)^5}
    & = 2 \frac{v^0 v_{\mu_1}v_{\mu_2}}{(v\cdot v)^4}
    - \frac{1}{3}\frac{\delta_{\mu_1}^0 v_{\mu_2} + \delta_{\mu_2}^0 v_{\mu_1} + v^0 g_{\mu_1\mu_2}}{(v\cdot v)^3}\,.
    \label{eq:original-integral}
\end{align}
By letting $v\cdot v = 1$ we have, for any tensor $N_{\mu\nu}$,
\begin{align}
\frac{1}{4\pi}\int d\Omega \,\frac{n^{\mu}n^{\nu}N_{\mu\nu}}{(n\cdot v)^5}
    & = 2 v^0 v^{\mu}v^{\nu}N_{\mu\nu}   - \frac{1}{3}(N_{0\nu}v^\nu + v^{\mu}N_{\mu 0} + v^0{N_\mu}^\mu)\,.
\label{eq:int-for-no-a}
\end{align}
Next, 
by differentiating \cref{eq:original-integral} with respect to $\tau$ and letting
$v\cdot v=1$ and $v\cdot a = 0$, we obtain
\begin{align}
    \frac{1}{4\pi}\int d\Omega\,\frac{n\cdot a}{(n\cdot v)^6}n_\mu N^{\mu\nu}n_\nu
& = -\frac{2}{5}\left[ a^0 v^\mu v^\nu N_{\mu\nu} + v^0 (a^\mu v^\nu + v^\mu a^\nu) N_{\mu\nu} \right]
+ \frac{1}{15}\left(  a^\mu N_{\mu 0} + N_{0\nu}a^\nu + a^0 {N^\mu}_\mu\right)\,. \label{eq:int-for-onea}
\end{align}
Next, by differentiating \cref{eq:original-integral} twice with respect to $\tau$, equating
the contribution containing $\dot{a}_\mu$ and that not containing $\dot{a}_\mu$ on both sides, and
then letting $v\cdot v=1$ and $v\cdot a=0$, we find
\begin{align}
\int \frac{d\Omega}{4\pi}
\frac{(n\cdot a)^2}{(n\cdot v)^7}
n^\mu n^\nu N_{\mu\nu}
& = (a\cdot a)\left[ -\frac{8}{15}v^0 v^{\mu}v^{\nu}N_{\mu\nu}  + \frac{1}{15}(N_{0\nu}v^\nu + v^{\mu}N_{\mu 0} + v^0{N_\mu}^\mu)\right] \notag \notag \\
& \quad+ \frac{2}{15}[a^0 a^\mu v^\nu +a^0 v^\mu a^\nu + v^0 a^\mu a^\nu]N_{\mu\nu}\,,\label{eq:int-for-asquared}\\
\int \frac{d\Omega}{4\pi}
\frac{n\cdot \dot{a}}{(n\cdot v)^6}
n^\mu n^\nu N_{\mu\nu}
& = (v\cdot \dot{a})\left[ \frac{16}{5}v^0 v^{\mu}v^{\nu}N_{\mu\nu}  - \frac{2}{5}(N_{0\nu}v^\nu + v^{\mu}N_{\mu 0} + v^0{N_\mu}^\mu)\right] \notag \\
& \quad - \frac{2}{5} [\dot{a}^0 v^{\mu}v^{\nu} + v^0\dot{a}^\mu v^\nu + v^0 v^\mu \dot{a}^\nu] N_{\mu\nu}   
+\frac{1}{15}(N^{0\nu}\dot{a}_\nu + \dot{a}_{\mu}N^{\mu 0} + \dot{a}^0{N_\mu}^\mu)\,. \label{eq:int-for-adot}
\end{align}

Let us denote the angular integrals with the measure
$d\Omega/4\pi$ of the first, second, third and fourth 
terms in \cref{eq:the-Y} by $I_1$, $I_2$, $I_3$
and $I_4$, respectively. These are performed by
using eqs.~\eqref{eq:int-for-asquared},
\eqref{eq:int-for-onea}, \eqref{eq:int-for-adot}
and \eqref{eq:int-for-no-a}, respectively.
Then, after using \cref{eq:formula-for-acceleration}
to express $F^{\mu\nu}v_\nu$ and $\dot{F}^{\mu\nu}v_\nu$ in terms of $a^\mu$ and $\dot{a}^\mu$, we find
\begin{align}
I_1 & = -3(a\cdot a)\left[ -\frac{8}{15}v^0(a\cdot a) 
    +\frac{2e}{15m}F^{0\mu}a_\mu + \frac{e^2}{15m^2}v^0 F_{\mu\nu}F^{\mu\nu}\right] +\frac{2e^2}{5m^2}v^0 a^\mu F_{\mu\nu}F^{\nu\lambda}a_\lambda\,,\\
I_2 & = 6\left[ - \frac{2}{5}a^0a^\mu\dot{a}_\mu 
    + \frac{2e}{5m}v^0 F^{\mu\nu}a_\nu \left(\dot{a}_\mu + \frac{e}{m}F_{\mu\lambda}a^\lambda\right)
 -\frac{e^2}{15m^2}F^{0\lambda}\dot{F}_{\lambda\mu}a^\mu
    - \frac{e^2}{15m^2}\dot{F}^{0\lambda}F_{\lambda \mu}a^\mu + \frac{e^2}{15m^2}a^0 F^{\mu\nu}\dot{F}_{\mu\nu} \right]\,,\\
I_3 & = -2(a\cdot a)\left[ -\frac{16}{5}v^0(a\cdot a) 
    +\frac{4e}{5m}F^{0\mu}a_\mu + \frac{2e^2}{5m^2}v^0 F_{\mu\nu}F^{\mu\nu}\right]
    +\frac{4}{5}\dot{a}^0(a\cdot a) -\frac{8e}{5m}v^0 a_\nu F^{\nu\mu}\dot{a}_\mu \notag \\
    & \quad + \frac{4e^2}{15m^2}F^{0\mu}F_{\mu\nu}\dot{a}^\nu  - \frac{2e^2}{15m^2}\dot{a}^0 F^{\mu\nu}F_{\mu\nu}\,,\\
 I_4 & = -2v^0\left(\dot{a}^\mu + \frac{2e}{m}F^{\mu\nu}a_\nu\right)
    \left(\dot{a}_\mu + \frac{2e}{m}F_{\mu\lambda}a^\lambda\right)
+ \frac{2e}{3m}\left( \dot{F}^{0\lambda} - \frac{e}{m}F^{0\alpha}{F_{\alpha}}^\lambda\right)
\left( \dot{a}_\lambda + \frac{2e}{m}F_{\lambda\beta}a^\beta\right)\notag \\ 
& \quad + \frac{e^2}{3m^2}v^0\left(\dot{F}^{\mu\nu} + \frac{e}{m}F^{\mu\lambda}{F_{\lambda}}^\nu\right)
\left(\dot{F}_{\mu\nu} + \frac{e}{m}F_{\mu\beta}{F^\beta}_\nu\right)\,.
\end{align}
For $I_3$ we used the equality $v\cdot \dot{a}=-a\cdot a$.
We can write the sum $I=I_1+I_2+I_3+I_4$ as
\begin{align}
    I = K_0 + \frac{e}{m}K_1 + \frac{e^2}{m^2}K_2 - \frac{4e^3}{3m^3}
    F^{0\alpha}F_{\alpha\beta}F^{\beta\nu}a_\nu 
    + \frac{e^4}{3m^4}v^0F_{\mu\nu}F^{\nu\lambda}F_{\lambda\alpha}F^{\alpha\mu}\,,
\end{align}
where
\begin{align}
    K_0 & = 2v^0[4(a\cdot a)^2 - \dot{a}\cdot \dot{a}] - \frac{12}{5}a^0(a\cdot \dot{a}) 
    + \frac{4}{5}\dot{a}^0(a\cdot a)\,,\label{eq:K0-expression}\\
    K_1 & = -4v^0 \dot{a}^\mu F_{\mu\nu}a^\nu 
    - 2(a\cdot a)F^{0\mu}a_\mu + \frac{2}{3}\dot{F}^{0\mu}\dot{a}_\mu\,,\\
    K_2 & = 6v^0 a_\mu F^{\mu\nu}F_{\nu\lambda}a^\lambda
    - v^0 (a\cdot a)F_{\mu\nu}F^{\mu\nu} - \frac{2}{5}F^{0\mu}F_{\mu\nu}\dot{a}^\nu
    + \frac{14}{15}\dot{F}^{0\lambda}F_{\lambda\mu}a^\mu \notag \\
    & \quad + \frac{1}{3}v^0 \dot{F}^{\mu\nu}\dot{F}_{\mu\nu}
    - \frac{2}{5}F^{0\lambda}\dot{F}_{\lambda\mu}a^\mu
    + \frac{2}{5}a^0 F^{\mu\nu}\dot{F}_{\mu\nu} - \frac{2}{15}\dot{a}^0 F^{\mu\nu}F_{\mu\nu}\,. 
\end{align}
These can be given in terms of the electric and magnetic
fields as
\begin{align}
    K_1 & = -4v^0[\dot{a}^0 \mathbf{a}\cdot\mathbf{E} - \dot{\mathbf{a}}\cdot(a^0\mathbf{E} + \mathbf{a}\times\mathbf{B})] 
    -2(a\cdot a)\mathbf{a}\cdot\mathbf{E} + \frac{2}{3}\dot{\mathbf{E}}\cdot\dot{\mathbf{a}}\,,\label{eq:K1-expression}\\
    K_2 & = 6v^0\left[ \|a^0\mathbf{E} + \mathbf{a}\times\mathbf{B}\|^2 - (\mathbf{a}\cdot\mathbf{E})^2\right]
    + 2v^0(a\cdot a)(\mathbf{E}^2 - \mathbf{B}^2) - \frac{2}{5}\mathbf{E}\cdot(\dot{a}^0\mathbf{E}+\dot{\mathbf{a}}\times\mathbf{B})
    + \frac{14}{15}\dot{\mathbf{E}}\cdot(a^0\mathbf{E} + \mathbf{a}\times\mathbf{B})\notag \\
& \quad - \frac{2}{3}v^0(\dot{\mathbf{E}}^2 - \dot{\mathbf{B}}^2) - \frac{2}{5}\mathbf{E}\cdot(a^0\dot{\mathbf{E}} + \mathbf{a}\times\dot{\mathbf{B}})
    - \frac{4}{5}a^0(\mathbf{E}\cdot{\dot{\mathbf{E}}} - \mathbf{B}\cdot\dot{\mathbf{B}})
    + \frac{4}{15}\dot{a}^0(\mathbf{E}^2 - \mathbf{B}^2)\,,
    \label{eq:K2-expression}
\end{align}
where we have used
\begin{align}
    F^{\mu\nu}a_\nu & = (\mathbf{a}\cdot\mathbf{E},a^0\mathbf{E} + \mathbf{a}\times\mathbf{B})\,,
\end{align}
and
\begin{align}
     F^{0\alpha}F_{\alpha\beta}F^{\beta\nu}a_\nu
     & = (\mathbf{a}\cdot\mathbf{E})(\mathbf{E}^2 - \mathbf{B}^2) + (\mathbf{a}\cdot\mathbf{B})(\mathbf{E}\cdot\mathbf{B})\,,\\
     F_{\mu\nu}F^{\nu\lambda}F_{\lambda\alpha}F^{\alpha\mu} & = 2(\mathbf{E}^2 -\mathbf{B}^2)^2 + 4(\mathbf{E}\cdot\mathbf{B})^2\,.
\end{align}
The energy emitted is given by
\begin{align}
 \braket{E} & = \frac{\hbar^2g_{ae}^2 }{16\pi m^2}
 \int I\,d\tau\,. \label{eq:energy-emitted-I}
\end{align}
\end{widetext}

\subsection{One-dimensional motion}

Let $\mathbf{B}=0$ and let $\mathbf{E}$ and $\mathbf{v}$ be in the $z$-direction. 
With the notation $v^z = w$, and with the prime denoting the $t$-derivative, we find from \cref{eq:K0-expression,eq:K1-expression,eq:K2-expression},
\newline
\newline
\begin{align}
\begin{split}
&K_0  = v^0\left[ \frac{26}{5}(w')^4 + \frac{8}{5}w(w')^2w'' + 2(1+w^2)(w'')^2\right]\,,\\
&\frac{e}{m} K_1  = v^0\left[ 2(w')^4 - \frac{2}{3}(1+w^2)(w'')^2- \frac{2}{3}w(w')^2w''\right]\,, \\
& \frac{e^2}{m^2}K_2 \\
    &= v^0\left[- \frac{122}{15}(w')^4 - \frac{2}{5}ww''(w')^2 - \frac{2}{3}(1+w^2)(w'')^2\right]\,,
\end{split}
\end{align}
 where we have used  $(e/m)E_z = -w'$, and
 \begin{align}
 \begin{split}
&-\frac{4e^3}{3m^3}
    F^{0\alpha}F_{\alpha\beta}F^{\beta\nu}a_\nu 
    + \frac{e^4}{3m^4}v^0F_{\mu\nu}F^{\nu\lambda}F_{\lambda\alpha}F^{\alpha\mu} \\
    & =  2v^0 (w')^4\,,
     \end{split}
\end{align}
Then,
\begin{align}
    I & = v^0\left[ \frac{16}{15}(w')^4 + \frac{2}{3}(1+w^2)(w'')^2 + \frac{8}{15}ww''(w')^2\right]\,.
\end{align}
For $w=2a_0\sin\omega_0 t$, we find the energy emitted
in one cycle from \cref{eq:energy-emitted-I} as
\begin{align}
E_{\textrm{linear}}^{(1)} = \frac{g_{ae}^2\hbar^2}{6m^2}(7a_0^2+1)a_0^2\omega_0^3\,. \label{eq:energy-emitted-linear}
\end{align}

\subsection{Circular motion with constant magnetic field}

If the magnetic field $\mathbf{B}$ 
is constant, $\mathbf{E}=\mathbf{0}$, and if the motion is stationary, then
\begin{align}
K_1 & = 4v^0\dot{\mathbf{a}}\cdot (\mathbf{a}\times \mathbf{B})\,,\\
K_2 & = 6v^0\|\mathbf{a}\times \mathbf{B}\|^2 - 2v^0(a\cdot a)\mathbf{B}^2\,,
\end{align}
so that
\begin{align}
I & = 2v^0[4(a\cdot a)^2 - \dot{a}\cdot \dot{a}] + 
\frac{4e}{m}v^0\dot{\mathbf{a}}\cdot (\mathbf{a}\times \mathbf{B})\notag \\
& \quad + \frac{e^2}{m^2}[6v^0\|\mathbf{a}\times \mathbf{B}\|^2 - 2v^0(a\cdot a)\mathbf{B}^2] \notag \\
& \quad +\frac{2e^4}{3m^4}v^0(\mathbf{B}^2)^2\,.
\end{align}
($a\cdot \dot{a}=0$ because $a\cdot a$ is constant in this case.) From \cref{eq:acceleration-time,eq:acceleration-space}, we find $a^0=0$ and
\begin{align}
    \mathbf{a} & =-\frac{e}{m}\mathbf{v}\times\mathbf{B}\,,\label{eq:accel-B-only}\\
    \dot{\mathbf{a}} & = - \frac{e}{m}\mathbf{a}\times\mathbf{B}\,.
\end{align}
Using the fact that $\mathbf{a}\perp \mathbf{B}$, we find
\begin{align}
    I = 2v^0\left[4(\mathbf{a}^2)^2 + \frac{3e^2}{m^2}\mathbf{B}^2\mathbf{a}^2 + \frac{e^4}{3m^4}(\mathbf{B}^2)^2 \right]\,. \label{eq:constant-B-case}
\end{align}

Now, suppose that the motion is circular with angular
frequency $\omega_0$.  We choose the magnetic field
to be in the $y$-direction and define
\begin{align}
a_0 = \frac{eB_y}{m\omega_0}\,.
\end{align}
Then, the motion of the electron can be described by
\begin{align}
z & = \frac{\sqrt{1-a_0^{-2}}}{\omega_0}\cos\omega_0 t\,, 
\label{eq:z-coord-const-B}\\
x & = \frac{\sqrt{1-a_0^{-2}}}{\omega_0}\sin\omega_0 t\,.
\label{eq:x-coord-const-B}
\end{align}
Then,
\begin{align}
    \mathbf{a}^2 & = a_0^2(a_0^2-1)\omega_0^2\,,
\end{align}
and
\begin{align}
    I & = v^0 \left(  8a_0^8 - 10a_0^6 + \frac{8}{3}a_0^4\right)\omega_0^4\,.
\end{align}
Thus, the energy of the axions emitted in one cycle is
\begin{align}
    E^{(1)}_{\textrm{circular}} & = \frac{g_{ae}^2\hbar}{4 m^2}\left( 4a_0^8 - 5a_0^6 + \frac{4}{3}a_0^4\right)\omega_0^3\,. \label{eq:power-for-B-const}
\end{align}
Notice that this is much larger than the energy emitted
in the linear case, \cref{eq:energy-emitted-linear}
if $a_0\gg 1$, i.e., if the electron is highly 
relativistic.

\section*{Appendix V: The emission spectrum for the constant magnetic field}
\label{appV}
Here we investigate the axion-emission
spectrum for a circular motion with a constant magnetic
field given by \cref{eq:z-coord-const-B,eq:x-coord-const-B}.

The motion is periodic, and the emission amplitude
can be expressed as a Fourier series with period $2\pi/\omega_0$. The (infinite) 
emission probability is formally of the form
\begin{align}
    P = \int_0^\infty dk \int d\Omega g(k,\Omega)
    \left| \int_{-\infty}^\infty \sum_{n=-\infty}^\infty c_n e^{i(k-n\omega_0)\xi}d\xi\right|^2\,,
\end{align}
where $\xi$ is a variable we identify later and
has the same period $2\pi/\omega_0$ as the time $t$.
We formally find
\begin{align}
P & = 2\pi \int_0^\infty dk \int d\Omega g(k,\Omega)
    \sum_{n=-\infty}^\infty |c_n|^2 \delta(k-n\omega_0) \notag\\
    &  \quad \times 2\pi\delta(0)\,, \label{eq:fourier-emission}
\end{align}
where $2\pi\delta(0)$ is formally the integral over $\xi$ from $-\infty$ to $\infty$.
Integration over $\xi$ from $-T/2$ to $T/2$ is equivalent to that over $t$ for the same
period if $T/2$ is a multiple of their period 
$2\pi/\omega_0$.  Thus, we can regard
$2\pi\delta(0)$ as the time duration $T$.  Hence, the emission rate, $R=P/T$, is
\begin{align}
    R & = 2\pi\sum_{n=1}^\infty \int d\Omega\, 
    g(n\omega_0,\Omega) |c_n|^2\,.
\end{align}
To use this formula to find the emission rate, we
need to find the Fourier coefficients $c_n$.

The periodicity of the motion also lets us use the first line of \cref{eq:amplitude-supp}, which
is the form of the amplitude before the integration by parts to remove the
contribution from the ``axion cloud'' around the electron.


\subsection{The spin-non-flip case}

The amplitude in the first line of \cref{eq:amplitude-supp} in the spin-non-flip case with the
spin in the positive or negative $y$-direction reads
\begin{align}
    \mathcal{A}_{(\mathbf{p},\mathbf{k},\alpha,\beta)}
    = \frac{g_{ae}}{2m}k_y \int e^{ik\cdot x(\tau)}d\tau\,.
\end{align}
Then, the (infinite) emission probability is
\begin{align}
    P_{\mathrm{em}}^{\mathrm{nf}} =\frac{\hbar g_{ae}^2}{4m^2}
    \int \frac{d^3\mathbf{k}}{(2\pi)^32k_0}k_y^2
    \left|\frac{1}{a_0}\int \,e^{ik\xi}\frac{dt}{d\xi}d\xi\right|^2\,,
    \label{eq:emission-non-f}
\end{align}
where we have used $d\tau = a_0 dt$.  The variable $\xi$ is defined by
$k\xi = k\cdot x = k t - \mathbf{k}\cdot\mathbf{x}$, where $k=k_0$. We define
the spherical polar coordinates $\theta$ and $\varphi$ by
\begin{align}
    (k_z,k_x,k_y) = (k\sin\theta\cos\varphi,k\sin\theta\sin\varphi,k\cos\theta)\,.
\end{align}
Using this and \cref{eq:z-coord-const-B,eq:x-coord-const-B},
we find
\begin{align}
    \xi = t - \frac{\sqrt{1-a_0^{-2}}}{\omega_0}\sin\theta \cos(\omega_0 t-\varphi)\,.
\end{align}
Comparing \cref{eq:fourier-emission,eq:emission-non-f}, 
we can make the following identification:
\begin{align}
    \int_0^\infty dk \int d\Omega\, g(k,\Omega)
    & = \frac{\hbar g_{ae}^2}{4m^2}\int \frac{d^3\mathbf{k}}{(2\pi)^32k_0}k_y^2\,,\\
    \frac{1}{a_0}\frac{dt}{d\xi} & = \sum_{n=-\infty}^\infty c_n e^{-in\omega_0\xi}\,,
\end{align}
where
\begin{align}
    \frac{dt}{d\xi} = \frac{1}{1+\sqrt{1-a_0^{-2}}\sin\theta\sin(\omega_0 t - \varphi)}\,, \label{eq:dt-dxi}
\end{align}
is a periodic function of $\xi$ with period $2\pi/\omega_0$. The Fourier coefficients
$c_n$ can be found as
\begin{align}
    c_n & = \frac{\omega_0}{2\pi a_0}\int_{0}^{2\pi/\omega_0} e^{in\omega_0\xi} \frac{dt}{d\xi} d\xi \notag\\
    & = \frac{1}{2\pi a_0}\int_0^{2\pi} e^{in[s - \beta\sin\theta \cos(s - \varphi)]} ds  \ \ (s=\omega_0t)\,,
\end{align} 
where $\beta = \sqrt{1- a_0^{-2}}$.
Then, by the formula
\begin{align}
    \frac{1}{2\pi}\int_0^{2\pi} e^{ins-z\cos s}ds
    = e^{-in\pi/2}J_n(z)\,,\label{eq:GR-formula}
\end{align}
which follows from eqs.~3.715.13 and 3.715.18 of Ref.~\cite{gradshteyn2014table},
where $J_n(x)$ is the Bessel function of order $n$,
we find
\begin{align}
c_n    & = \frac{e^{in\varphi-in\pi/2}}{a_0}J_n(n\sqrt{1-a_0^{-2}}\sin\theta)\,.
\end{align}
Then, the rate of emission is
\begin{widetext}
\begin{align}
    R_{\mathrm{em}}^{\mathrm{nf}} & = \frac{\hbar g_{ae}^2\omega_0^3}{16\pi m^2}\sum_{n=1}^\infty
    \frac{n^3}{a_0^2}\int_{0}^\pi \left|J_n\left(n\sqrt{1-a_0^{-2}}\sin\theta\right)\right|^2\cos^2\theta\sin\theta\,d\theta\,,
    \label{eq:rate-with-Fourier}
\end{align}
and the power is
\begin{align}
     S_{\mathrm{em}}^{\mathrm{nf}} & = \frac{\hbar^2 g_{ae}^2\omega_0^4}{16\pi m^2}\sum_{n=1}^\infty
    \frac{n^4}{a_0^2}\int_{0}^\pi \left|J_n\left(n\sqrt{1-a_0^{-2}}\sin\theta\right)\right|^2\cos^2\theta\sin\theta\,d\theta\,.
    \label{eq:power-with-Fourier}
\end{align}
It is possible to carry out the summation and $\theta$-integral in \cref{eq:power-with-Fourier}.  Define $u=s-\alpha \cos s$, where 
$0 < \alpha < 1$.  then, the variable $du/ds > 0$.  We write \cref{eq:GR-formula}
with $z=n\alpha$ as
\begin{align}
    e^{-in\pi/2}J_n(n\alpha) = \frac{1}{2\pi}\int_0^{2\pi}
    \frac{e^{inu}}{1+\alpha\sin s}\,du\,.
\end{align}
Then,
\begin{align}
    e^{-in\pi/2}n^2 J_n(n\alpha) = - \frac{1}{2\pi}\int_0^{2\pi}
    \frac{d^2\ }{du^2}\left( \frac{1}{1+\alpha\sin s}\right)e^{inu} du\,.
\end{align}
Since $|J_n(n\alpha)|^2 = |J_{-n}(-n\alpha)|^2$, Parceval's theorem implies
\begin{align}
    \sum_{n=1}^\infty n^4|J_n(n\alpha)|^2 & = \frac{1}{4\pi}\int_0^{2\pi}
    \left|\frac{d^2\ }{du^2}\left(\frac{1}{1+\alpha\sin s}\right)\right|^2\,du\notag \\
    & = \frac{(27\alpha^6 + 472\alpha^4+592\alpha^2+64)\alpha^2}{256(1-\alpha^2)^{13/2}}\,.
\end{align}
Then, we perform the $\theta$-integral
with $\alpha=\sqrt{1-a_0^{-2}}\sin\theta$.  We find the energy emitted in one cycle as
\begin{align}
    E_{\textrm{circular}}^{(1)\mathrm{nf}} 
    = \frac{g_{ae}^2\hbar^2}{8m^2}\left(\frac{1}{3}a_0^8-\frac{3}{5}a_0^6+\frac{4}{15}a_0^4\right)\omega_0^3\,.
\end{align}

\subsection{The spin-flip case}

For the spin-flip case, the amplitude in the first line of \cref{eq:amplitude-supp} reads
\begin{align}
    \mathcal{A}_{(\mathbf{p},\mathbf{k},\mp,\pm)} = 
    \frac{g_{ae}}{2m} \int \left[ k(n_z\pm in_x) - \frac{k\cdot p + mk}{m(p_0 + m)}(p_z\pm ip_x)\right]e^{\mp i\omega_0 t}e^{ik\cdot x}\,d\tau\,.
\end{align}
Since $n_z \pm in_x = \sin\theta\,e^{\pm i\varphi}$ and 
$p_z\pm i p_x = m\sqrt{a_0^2-1}(-\sin\omega_0 t \pm i\cos\omega_0 t) = \pm i m\sqrt{a_0^2-1}e^{\pm i\omega_0 t}$,
we find
\begin{align}
    \mathcal{A}_{(\mathbf{p},\mathbf{k},\mp,\pm)} = \frac{g_{ae}}{2m} \int \left[ k\sin\theta e^{\mp i (\omega_0 t- \varphi)} \mp i \sqrt{a_0^2-1}\frac{k\cdot p + mk}{p_0 + m}\right]e^{ik\cdot x}\,d\tau\,.
\end{align}
In our case, $p_0 = ma_0$ is constant, and
\begin{align}
    \int k\cdot p\,e^{ik\cdot x}d\tau = -i m \int \frac{d\ }{d\tau}e^{ik\cdot x}\,d\tau = 0\,, 
\end{align}
by periodicity.  Thus, we find
\begin{align}
    \mathcal{A}_{(\mathbf{p},\mathbf{k},\mp,\pm)} 
    & = \mp i\frac{g_{ae}k}{2m} \int \left[ \sqrt{\frac{a_0-1}{a_0+1}}
    \pm i \sin\theta\,e^{\mp i(\omega_0 t-\varphi)}\right]e^{ik\cdot x}\,d\tau\,.
\end{align}
Then, we can make the following identification in \cref{eq:fourier-emission}:
\begin{align}
     \int_0^\infty dk \int d\Omega g(k,\Omega)
     = \frac{\hbar g_{ae}^2}{4m^2}\int \frac{d^3\mathbf{k}}{(2\pi)^32k}k^2\,,
\end{align}
and
\begin{align}
     \left[\sqrt{\frac{a_0-1}{a_0+1}} \pm i \sin \theta e^{\mp i(\omega_0 t-\varphi)}\right]\frac{dt}{d\xi}  = \sum_{n=-\infty}^\infty c_n^{\pm} e^{-in\omega_0\xi}\,,
\end{align}
where $dt/d\xi$ is given by \cref{eq:dt-dxi}.

Proceeding as in the spin-non-flip case, we find
\begin{align}
c_n^{\pm} 
& = \frac{e^{in\varphi}}{2\pi a_0}\int_0^{2\pi} \left[ \sqrt{\frac{a_0-1}{a_0+1}}e^{in(s - \beta\sin\theta \cos s)}
\pm  i\sin\theta e^{i(n\mp 1)s - \beta\sin\theta\cos s}\right]ds\,.
\end{align}
This integral is evaluated using \cref{eq:GR-formula} again, and we obtain
\begin{align}
   a_0c_n^{\pm}e^{-in\varphi} & = e^{-i\frac{n\pi}{2}}\sqrt{\frac{a_0-1}{a_0+1}} J_n\left(n\sqrt{1-a_0^{-2}}
    \sin\theta\right) \pm  i\sin\theta e^{-i(n\mp 1)\pi/2} J_{n\mp 1}\left(n\sqrt{1-a_0^{-2}}\sin\theta\right)\,.
\end{align}
Then, we obtain the rate of emission as
\begin{align}
    R_{\mathrm{em}}^{\mathrm{f},\pm} 
    & = \frac{\hbar g_{ae}^2\omega_0^3}{16\pi m^2}\sum_{n=1}^\infty
    \frac{n^3}{a_0^2}\int_{0}^\pi \left|\sqrt{\frac{a_0-1}{a_0+1}} J_n\left(n\sqrt{1-a_0^{-2}}
    \sin\theta\right) - \sin\theta  J_{n\mp 1}\left(n\sqrt{1-a_0^{-2}}\sin\theta\right)\right|^2\sin\theta\,d\theta\,,
    \label{eq:rate-for-spin-flip}
\end{align}
and the power is
\begin{align}
      S_{\mathrm{em}}^{\mathrm{f},\pm}
    & = \frac{\hbar^2 g_{ae}^2\omega_0^4}{16\pi m^2}\sum_{n=1}^\infty
    \frac{n^4}{a_0^2}\int_{0}^\pi \left|\sqrt{\frac{a_0-1}{a_0+1}} J_n\left(n\sqrt{1-a_0^{-2}}
    \sin\theta\right) - \sin\theta  J_{n\mp 1}\left(n\sqrt{1-a_0^{-2}}\sin\theta\right)\right|^2\sin\theta\,d\theta\,.
    \label{eq:power-for-spin-flip}
\end{align}

Parceval's theorem can be used to find the average of the up-to-down and down-to-up 
spin-flip powers of emission because, although
$|c_n^{(\pm)}|\neq |c_{-n}^{(\pm)}|$, we have $|c_{-n}^{(+)}| = |c_{n}^{(-)}|$.
That is,
\begin{align}
     \sum_{n=1}^\infty n^4(|c_n^{(+)}|^2 + |c_n^{(-)}|^2)
     = \sum_{n=-\infty}^\infty n^4|c_n^{(+)}|^2\,.
\end{align}
Thus, Parceval's theorem implies
\begin{align}
      Y:= \sum_{n=1}^\infty n^4(|c_n^{(+)}|^2 + |c_n^{(-)}|^2) 
    & = \frac{1}{4\pi}\int_0^{2\pi}\left| \frac{d^2\ }{du^2}\left\{ \frac{1}{1+\alpha\sin s}\left[ \sqrt{\frac{a_0-1}{a_0+1}} + i\sin\theta e^{-is}\right] \right\}\right|^2
    (1+\alpha \sin s)ds\,.
\end{align}
This integral can be evaluated 
as
\begin{align}
    Y & =\frac{(27\alpha^6 + 472\alpha^4 + 592\alpha^2+64)}{256(1-\alpha^2)^{13/2}}\left( \alpha^2\frac{a_0-1}{a_0+1} - 2\alpha\sqrt{\frac{a_0-1}{a_0+1}}\sin\theta\right)
    - \frac{45\alpha^8+560\alpha^6-480\alpha^4-1152\alpha^2 - 128}{256(1-\alpha^2)^{13/2}}\sin^2\theta\,.
\end{align}
For $\alpha = \sqrt{1-a_0^{-2}}\sin\theta$ we find
\begin{align}
    \alpha^2\frac{a_0-1}{a_0+1} - 2\alpha\sqrt{\frac{a_0-1}{a_0+1}}\sin\theta 
    = -\alpha^2\,.
\end{align}
Hence
\begin{align}
    Y & =-\frac{27\alpha^8 + 472\alpha^6 + 592\alpha^4+64\alpha^2}{256(1-\alpha^2)^{13/2}}
    - \frac{45\alpha^8+560\alpha^6-480\alpha^4-1152\alpha^2 - 128}{256(1-\alpha^2)^{13/2}}\sin^2\theta\,.
\end{align}
Then, by evaluating the angular integral in \cref{eq:power-for-spin-flip}, we find the spin-averaged axion energy emitted in one cycle as
\begin{align}
     E_{\textrm{circular}}^{(1)\mathbf{f,av}} & = \frac{\hbar^2 g_{ae}^2}{8 m^2}\left( \frac{23}{3}a_0^8 - \frac{47}{5}a_0^6 + \frac{12}{5}a_0^4\right)\omega_0^4\,.
\end{align}
We find $E_{\textrm{circular}}^{(1)\mathrm{nf}}+E_{\textrm{circular}}^{(1)\mathbf{f,av}}
= E_{\textrm{circular}}^{(1)}$, where $E_{\textrm{circular}}^{(1)}$ is given by \cref{eq:power-for-B-const}, as expected.

\end{widetext}
The ratio between power and rate (i.e. dividing \cref{eq:power-with-Fourier} by \cref{eq:rate-with-Fourier} and \cref{eq:power-for-spin-flip} by \cref{eq:rate-for-spin-flip}) can be used as a function of $a_0$ to find the typical axion energy. We find numerically that, in the spin-non-flip case, the ratio scales as $\sim 3 a_0^{3} \omega_0$. In the spin-flip case, the ratio scales as $\sim 3.7 a_0^3 \omega_0$ and $\sim 2 a_0^3 \omega_0$ for initial spin in the positive or negative $y$-direction, respectively. We also find numerically that the largest part of the emitted energy comes from the spin-flip case for initial spin in the positive $y$-direction. We can therefore take the typical axion energy to be $\sim 4 a_0^3 \omega_0$. Although this result holds for a constant magnetic field only, we expect the same to be qualitatively true for axion emission from an electron accelerated by two laser fields, as described in the main text.


\begin{thebibliography}{42}%
\makeatletter
\providecommand \@ifxundefined [1]{%
 \@ifx{#1\undefined}
}%
\providecommand \@ifnum [1]{%
 \ifnum #1\expandafter \@firstoftwo
 \else \expandafter \@secondoftwo
 \fi
}%
\providecommand \@ifx [1]{%
 \ifx #1\expandafter \@firstoftwo
 \else \expandafter \@secondoftwo
 \fi
}%
\providecommand \natexlab [1]{#1}%
\providecommand \enquote  [1]{``#1''}%
\providecommand \bibnamefont  [1]{#1}%
\providecommand \bibfnamefont [1]{#1}%
\providecommand \citenamefont [1]{#1}%
\providecommand \href@noop [0]{\@secondoftwo}%
\providecommand \href [0]{\begingroup \@sanitize@url \@href}%
\providecommand \@href[1]{\@@startlink{#1}\@@href}%
\providecommand \@@href[1]{\endgroup#1\@@endlink}%
\providecommand \@sanitize@url [0]{\catcode `\\12\catcode `\$12\catcode `\&12\catcode `\#12\catcode `\^12\catcode `\_12\catcode `\%12\relax}%
\providecommand \@@startlink[1]{}%
\providecommand \@@endlink[0]{}%
\providecommand \url  [0]{\begingroup\@sanitize@url \@url }%
\providecommand \@url [1]{\endgroup\@href {#1}{\urlprefix }}%
\providecommand \urlprefix  [0]{URL }%
\providecommand \Eprint [0]{\href }%
\providecommand \doibase [0]{https://doi.org/}%
\providecommand \selectlanguage [0]{\@gobble}%
\providecommand \bibinfo  [0]{\@secondoftwo}%
\providecommand \bibfield  [0]{\@secondoftwo}%
\providecommand \translation [1]{[#1]}%
\providecommand \BibitemOpen [0]{}%
\providecommand \bibitemStop [0]{}%
\providecommand \bibitemNoStop [0]{.\EOS\space}%
\providecommand \EOS [0]{\spacefactor3000\relax}%
\providecommand \BibitemShut  [1]{\csname bibitem#1\endcsname}%
\let\auto@bib@innerbib\@empty
\bibitem [{\citenamefont {Peccei}\ and\ \citenamefont {Quinn}(1977{\natexlab{a}})}]{Peccei:1977hh}%
  \BibitemOpen
  \bibfield  {author} {\bibinfo {author} {\bibfnamefont {R.~D.}\ \bibnamefont {Peccei}}\ and\ \bibinfo {author} {\bibfnamefont {H.~R.}\ \bibnamefont {Quinn}},\ }\bibfield  {title} {\bibinfo {title} {{CP Conservation in the Presence of Instantons}},\ }\href {https://doi.org/10.1103/PhysRevLett.38.1440} {\bibfield  {journal} {\bibinfo  {journal} {Phys. Rev. Lett.}\ }\textbf {\bibinfo {volume} {38}},\ \bibinfo {pages} {1440} (\bibinfo {year} {1977}{\natexlab{a}})}\BibitemShut {NoStop}%
\bibitem [{\citenamefont {Peccei}\ and\ \citenamefont {Quinn}(1977{\natexlab{b}})}]{Peccei:1977ur}%
  \BibitemOpen
  \bibfield  {author} {\bibinfo {author} {\bibfnamefont {R.~D.}\ \bibnamefont {Peccei}}\ and\ \bibinfo {author} {\bibfnamefont {H.~R.}\ \bibnamefont {Quinn}},\ }\bibfield  {title} {\bibinfo {title} {{Constraints Imposed by CP Conservation in the Presence of Instantons}},\ }\href {https://doi.org/10.1103/PhysRevD.16.1791} {\bibfield  {journal} {\bibinfo  {journal} {Phys. Rev. D}\ }\textbf {\bibinfo {volume} {16}},\ \bibinfo {pages} {1791} (\bibinfo {year} {1977}{\natexlab{b}})}\BibitemShut {NoStop}%
\bibitem [{\citenamefont {Weinberg}(1978)}]{Weinberg:1977ma}%
  \BibitemOpen
  \bibfield  {author} {\bibinfo {author} {\bibfnamefont {S.}~\bibnamefont {Weinberg}},\ }\bibfield  {title} {\bibinfo {title} {{A New Light Boson?}},\ }\href {https://doi.org/10.1103/PhysRevLett.40.223} {\bibfield  {journal} {\bibinfo  {journal} {Phys. Rev. Lett.}\ }\textbf {\bibinfo {volume} {40}},\ \bibinfo {pages} {223} (\bibinfo {year} {1978})}\BibitemShut {NoStop}%
\bibitem [{\citenamefont {Wilczek}(1978)}]{Wilczek:1977pj}%
  \BibitemOpen
  \bibfield  {author} {\bibinfo {author} {\bibfnamefont {F.}~\bibnamefont {Wilczek}},\ }\bibfield  {title} {\bibinfo {title} {{Problem of Strong $P$ and $T$ Invariance in the Presence of Instantons}},\ }\href {https://doi.org/10.1103/PhysRevLett.40.279} {\bibfield  {journal} {\bibinfo  {journal} {Phys. Rev. Lett.}\ }\textbf {\bibinfo {volume} {40}},\ \bibinfo {pages} {279} (\bibinfo {year} {1978})}\BibitemShut {NoStop}%
\bibitem [{\citenamefont {Preskill}\ \emph {et~al.}(1983)\citenamefont {Preskill}, \citenamefont {Wise},\ and\ \citenamefont {Wilczek}}]{Preskill:1982cy}%
  \BibitemOpen
  \bibfield  {author} {\bibinfo {author} {\bibfnamefont {J.}~\bibnamefont {Preskill}}, \bibinfo {author} {\bibfnamefont {M.~B.}\ \bibnamefont {Wise}},\ and\ \bibinfo {author} {\bibfnamefont {F.}~\bibnamefont {Wilczek}},\ }\bibfield  {title} {\bibinfo {title} {{Cosmology of the Invisible Axion}},\ }\href {https://doi.org/10.1016/0370-2693(83)90637-8} {\bibfield  {journal} {\bibinfo  {journal} {Phys. Lett. B}\ }\textbf {\bibinfo {volume} {120}},\ \bibinfo {pages} {127} (\bibinfo {year} {1983})}\BibitemShut {NoStop}%
\bibitem [{\citenamefont {Abbott}\ and\ \citenamefont {Sikivie}(1983)}]{Abbott:1982af}%
  \BibitemOpen
  \bibfield  {author} {\bibinfo {author} {\bibfnamefont {L.~F.}\ \bibnamefont {Abbott}}\ and\ \bibinfo {author} {\bibfnamefont {P.}~\bibnamefont {Sikivie}},\ }\bibfield  {title} {\bibinfo {title} {{A Cosmological Bound on the Invisible Axion}},\ }\href {https://doi.org/10.1016/0370-2693(83)90638-X} {\bibfield  {journal} {\bibinfo  {journal} {Phys. Lett. B}\ }\textbf {\bibinfo {volume} {120}},\ \bibinfo {pages} {133} (\bibinfo {year} {1983})}\BibitemShut {NoStop}%
\bibitem [{\citenamefont {Dine}\ and\ \citenamefont {Fischler}(1983)}]{Dine:1982ah}%
  \BibitemOpen
  \bibfield  {author} {\bibinfo {author} {\bibfnamefont {M.}~\bibnamefont {Dine}}\ and\ \bibinfo {author} {\bibfnamefont {W.}~\bibnamefont {Fischler}},\ }\bibfield  {title} {\bibinfo {title} {{The Not So Harmless Axion}},\ }\href {https://doi.org/10.1016/0370-2693(83)90639-1} {\bibfield  {journal} {\bibinfo  {journal} {Phys. Lett. B}\ }\textbf {\bibinfo {volume} {120}},\ \bibinfo {pages} {137} (\bibinfo {year} {1983})}\BibitemShut {NoStop}%
\bibitem [{\citenamefont {Anastassopoulos}\ \emph {et~al.}(2017)\citenamefont {Anastassopoulos} \emph {et~al.}}]{CAST:2017uph}%
  \BibitemOpen
  \bibfield  {author} {\bibinfo {author} {\bibfnamefont {V.}~\bibnamefont {Anastassopoulos}} \emph {et~al.} (\bibinfo {collaboration} {CAST}),\ }\bibfield  {title} {\bibinfo {title} {{New CAST Limit on the Axion-Photon Interaction}},\ }\href {https://doi.org/10.1038/nphys4109} {\bibfield  {journal} {\bibinfo  {journal} {Nature Phys.}\ }\textbf {\bibinfo {volume} {13}},\ \bibinfo {pages} {584} (\bibinfo {year} {2017})},\ \Eprint {https://arxiv.org/abs/1705.02290} {arXiv:1705.02290 [hep-ex]} \BibitemShut {NoStop}%
\bibitem [{\citenamefont {Ballou}\ \emph {et~al.}(2015)\citenamefont {Ballou} \emph {et~al.}}]{OSQAR:2015qdv}%
  \BibitemOpen
  \bibfield  {author} {\bibinfo {author} {\bibfnamefont {R.}~\bibnamefont {Ballou}} \emph {et~al.} (\bibinfo {collaboration} {OSQAR}),\ }\bibfield  {title} {\bibinfo {title} {{New exclusion limits on scalar and pseudoscalar axionlike particles from light shining through a wall}},\ }\href {https://doi.org/10.1103/PhysRevD.92.092002} {\bibfield  {journal} {\bibinfo  {journal} {Phys. Rev. D}\ }\textbf {\bibinfo {volume} {92}},\ \bibinfo {pages} {092002} (\bibinfo {year} {2015})},\ \Eprint {https://arxiv.org/abs/1506.08082} {arXiv:1506.08082 [hep-ex]} \BibitemShut {NoStop}%
\bibitem [{\citenamefont {Beyer}\ \emph {et~al.}(2020)\citenamefont {Beyer}, \citenamefont {Marocco}, \citenamefont {Bingham},\ and\ \citenamefont {Gregori}}]{Beyer:2020dag}%
  \BibitemOpen
  \bibfield  {author} {\bibinfo {author} {\bibfnamefont {K.~A.}\ \bibnamefont {Beyer}}, \bibinfo {author} {\bibfnamefont {G.}~\bibnamefont {Marocco}}, \bibinfo {author} {\bibfnamefont {R.}~\bibnamefont {Bingham}},\ and\ \bibinfo {author} {\bibfnamefont {G.}~\bibnamefont {Gregori}},\ }\bibfield  {title} {\bibinfo {title} {{Axion detection through resonant photon-photon collisions}},\ }\href {https://doi.org/10.1103/PhysRevD.101.095018} {\bibfield  {journal} {\bibinfo  {journal} {Phys. Rev. D}\ }\textbf {\bibinfo {volume} {101}},\ \bibinfo {pages} {095018} (\bibinfo {year} {2020})},\ \Eprint {https://arxiv.org/abs/2001.03392} {arXiv:2001.03392 [hep-ph]} \BibitemShut {NoStop}%
\bibitem [{\citenamefont {Beyer}\ \emph {et~al.}(2022)\citenamefont {Beyer}, \citenamefont {Marocco}, \citenamefont {Bingham},\ and\ \citenamefont {Gregori}}]{Beyer:2021mzq}%
  \BibitemOpen
  \bibfield  {author} {\bibinfo {author} {\bibfnamefont {K.~A.}\ \bibnamefont {Beyer}}, \bibinfo {author} {\bibfnamefont {G.}~\bibnamefont {Marocco}}, \bibinfo {author} {\bibfnamefont {R.}~\bibnamefont {Bingham}},\ and\ \bibinfo {author} {\bibfnamefont {G.}~\bibnamefont {Gregori}},\ }\bibfield  {title} {\bibinfo {title} {{Light-shining-through-wall axion detection experiments with a stimulating laser}},\ }\href {https://doi.org/10.1103/PhysRevD.105.035031} {\bibfield  {journal} {\bibinfo  {journal} {Phys. Rev. D}\ }\textbf {\bibinfo {volume} {105}},\ \bibinfo {pages} {035031} (\bibinfo {year} {2022})},\ \Eprint {https://arxiv.org/abs/2109.14663} {arXiv:2109.14663 [hep-ph]} \BibitemShut {NoStop}%
\bibitem [{\citenamefont {Sikivie}(2021)}]{RevModPhys.93.015004}%
  \BibitemOpen
  \bibfield  {author} {\bibinfo {author} {\bibfnamefont {P.}~\bibnamefont {Sikivie}},\ }\bibfield  {title} {\bibinfo {title} {Invisible axion search methods},\ }\href {https://doi.org/10.1103/RevModPhys.93.015004} {\bibfield  {journal} {\bibinfo  {journal} {Rev. Mod. Phys.}\ }\textbf {\bibinfo {volume} {93}},\ \bibinfo {pages} {015004} (\bibinfo {year} {2021})}\BibitemShut {NoStop}%
\bibitem [{\citenamefont {Dent}\ \emph {et~al.}(2020)\citenamefont {Dent}, \citenamefont {Dutta}, \citenamefont {Kim}, \citenamefont {Liao}, \citenamefont {Mahapatra}, \citenamefont {Sinha},\ and\ \citenamefont {Thompson}}]{PhysRevLett.124.211804}%
  \BibitemOpen
  \bibfield  {author} {\bibinfo {author} {\bibfnamefont {J.~B.}\ \bibnamefont {Dent}}, \bibinfo {author} {\bibfnamefont {B.}~\bibnamefont {Dutta}}, \bibinfo {author} {\bibfnamefont {D.}~\bibnamefont {Kim}}, \bibinfo {author} {\bibfnamefont {S.}~\bibnamefont {Liao}}, \bibinfo {author} {\bibfnamefont {R.}~\bibnamefont {Mahapatra}}, \bibinfo {author} {\bibfnamefont {K.}~\bibnamefont {Sinha}},\ and\ \bibinfo {author} {\bibfnamefont {A.}~\bibnamefont {Thompson}},\ }\bibfield  {title} {\bibinfo {title} {New directions for axion searches via scattering at reactor neutrino experiments},\ }\href {https://doi.org/10.1103/PhysRevLett.124.211804} {\bibfield  {journal} {\bibinfo  {journal} {Phys. Rev. Lett.}\ }\textbf {\bibinfo {volume} {124}},\ \bibinfo {pages} {211804} (\bibinfo {year} {2020})}\BibitemShut {NoStop}%
\bibitem [{\citenamefont {Piazza}\ \emph {et~al.}(2022)\citenamefont {Piazza}, \citenamefont {Willingale},\ and\ \citenamefont {Zuegel}}]{mp3}%
  \BibitemOpen
  \bibfield  {author} {\bibinfo {author} {\bibfnamefont {A.~D.}\ \bibnamefont {Piazza}}, \bibinfo {author} {\bibfnamefont {L.}~\bibnamefont {Willingale}},\ and\ \bibinfo {author} {\bibfnamefont {J.~D.}\ \bibnamefont {Zuegel}},\ }\href {https://arxiv.org/abs/2211.13187} {\bibinfo {title} {Multi-petawatt physics prioritization (mp3) workshop report}} (\bibinfo {year} {2022}),\ \Eprint {https://arxiv.org/abs/2211.13187} {arXiv:2211.13187 [hep-ph]} \BibitemShut {NoStop}%
\bibitem [{\citenamefont {Baĭer}\ and\ \citenamefont {Katkov}(1967)}]{BAIER1967492}%
  \BibitemOpen
  \bibfield  {author} {\bibinfo {author} {\bibfnamefont {V.}~\bibnamefont {Baĭer}}\ and\ \bibinfo {author} {\bibfnamefont {V.}~\bibnamefont {Katkov}},\ }\bibfield  {title} {\bibinfo {title} {Quantum effects in magnetic bremsstrahlung},\ }\href {https://doi.org/https://doi.org/10.1016/0375-9601(67)90003-5} {\bibfield  {journal} {\bibinfo  {journal} {Physics Letters A}\ }\textbf {\bibinfo {volume} {25}},\ \bibinfo {pages} {492} (\bibinfo {year} {1967})}\BibitemShut {NoStop}%
\bibitem [{\citenamefont {Baĭer}\ and\ \citenamefont {Katkov}(1968)}]{Baier_Katkov_2}%
  \BibitemOpen
  \bibfield  {author} {\bibinfo {author} {\bibfnamefont {V.}~\bibnamefont {Baĭer}}\ and\ \bibinfo {author} {\bibfnamefont {V.}~\bibnamefont {Katkov}},\ }\bibfield  {title} {\bibinfo {title} {Processes involved in the motion of high energy particles in a magnetic field},\ }\href@noop {} {\bibfield  {journal} {\bibinfo  {journal} {JETP}\ }\textbf {\bibinfo {volume} {26}},\ \bibinfo {pages} {854} (\bibinfo {year} {1968})}\BibitemShut {NoStop}%
\bibitem [{\citenamefont {Baĭer}\ and\ \citenamefont {Katkov}(1969)}]{Baier_Katkov_3}%
  \BibitemOpen
  \bibfield  {author} {\bibinfo {author} {\bibfnamefont {V.}~\bibnamefont {Baĭer}}\ and\ \bibinfo {author} {\bibfnamefont {V.}~\bibnamefont {Katkov}},\ }\bibfield  {title} {\bibinfo {title} {Quasiclassical theory of bremsstrahlung by relativistic particles},\ }\href@noop {} {\bibfield  {journal} {\bibinfo  {journal} {JETP}\ }\textbf {\bibinfo {volume} {28}},\ \bibinfo {pages} {807} (\bibinfo {year} {1969})}\BibitemShut {NoStop}%
\bibitem [{\citenamefont {Baĭer}(1972)}]{Baier_1972}%
  \BibitemOpen
  \bibfield  {author} {\bibinfo {author} {\bibfnamefont {V.~N.}\ \bibnamefont {Baĭer}},\ }\bibfield  {title} {\bibinfo {title} {Radiative polarization of electrons in storage rings},\ }\href {https://doi.org/10.1070/PU1972v014n06ABEH004751} {\bibfield  {journal} {\bibinfo  {journal} {Soviet Physics Uspekhi}\ }\textbf {\bibinfo {volume} {14}},\ \bibinfo {pages} {695} (\bibinfo {year} {1972})}\BibitemShut {NoStop}%
\bibitem [{\citenamefont {Bjorken}\ and\ \citenamefont {Drell}(1965)}]{BjorkenDrell}%
  \BibitemOpen
  \bibfield  {author} {\bibinfo {author} {\bibfnamefont {J.~D.}\ \bibnamefont {Bjorken}}\ and\ \bibinfo {author} {\bibfnamefont {S.~D.}\ \bibnamefont {Drell}},\ }\href@noop {} {\emph {\bibinfo {title} {{Relativistic Quantum Mechanics}}}},\ International Series In Pure and Applied Physics\ (\bibinfo  {publisher} {McGraw-Hill},\ \bibinfo {address} {New York},\ \bibinfo {year} {1965})\BibitemShut {NoStop}%
\bibitem [{\citenamefont {Thomas}(1926{\natexlab{a}})}]{Thomas:1926dy}%
  \BibitemOpen
  \bibfield  {author} {\bibinfo {author} {\bibfnamefont {L.~H.}\ \bibnamefont {Thomas}},\ }\bibfield  {title} {\bibinfo {title} {{The motion of a spinning electron}},\ }\href {https://doi.org/10.1038/117514a0} {\bibfield  {journal} {\bibinfo  {journal} {Nature}\ }\textbf {\bibinfo {volume} {117}},\ \bibinfo {pages} {514} (\bibinfo {year} {1926}{\natexlab{a}})}\BibitemShut {NoStop}%
\bibitem [{\citenamefont {Bargmann}\ \emph {et~al.}(1959)\citenamefont {Bargmann}, \citenamefont {Michel},\ and\ \citenamefont {Telegdi}}]{Bargmann:1959gz}%
  \BibitemOpen
  \bibfield  {author} {\bibinfo {author} {\bibfnamefont {V.}~\bibnamefont {Bargmann}}, \bibinfo {author} {\bibfnamefont {L.}~\bibnamefont {Michel}},\ and\ \bibinfo {author} {\bibfnamefont {V.~L.}\ \bibnamefont {Telegdi}},\ }\bibfield  {title} {\bibinfo {title} {{Precession of the polarization of particles moving in a homogeneous electromagnetic field}},\ }\href {https://doi.org/10.1103/PhysRevLett.2.435} {\bibfield  {journal} {\bibinfo  {journal} {Phys. Rev. Lett.}\ }\textbf {\bibinfo {volume} {2}},\ \bibinfo {pages} {435} (\bibinfo {year} {1959})}\BibitemShut {NoStop}%
\bibitem [{\citenamefont {Higuchi}\ and\ \citenamefont {Martin}(2006)}]{Higuchi:2005an}%
  \BibitemOpen
  \bibfield  {author} {\bibinfo {author} {\bibfnamefont {A.}~\bibnamefont {Higuchi}}\ and\ \bibinfo {author} {\bibfnamefont {G.~D.~R.}\ \bibnamefont {Martin}},\ }\bibfield  {title} {\bibinfo {title} {{Radiation reaction on charged particles in three-dimensional motion in classical and quantum electrodynamics}},\ }\href {https://doi.org/10.1103/PhysRevD.73.025019} {\bibfield  {journal} {\bibinfo  {journal} {Phys. Rev. D}\ }\textbf {\bibinfo {volume} {73}},\ \bibinfo {pages} {025019} (\bibinfo {year} {2006})},\ \Eprint {https://arxiv.org/abs/quant-ph/0510043} {arXiv:quant-ph/0510043} \BibitemShut {NoStop}%
\bibitem [{\citenamefont {Raffelt}(1996)}]{Raffelt:1996wa}%
  \BibitemOpen
  \bibfield  {author} {\bibinfo {author} {\bibfnamefont {G.~G.}\ \bibnamefont {Raffelt}},\ }\href@noop {} {\emph {\bibinfo {title} {Stars as laboratories for fundamental physics: The astrophysics of neutrinos, axions, and other weakly interacting particles}}}\ (\bibinfo  {publisher} {The University of Chicago Press},\ \bibinfo {year} {1996})\BibitemShut {NoStop}%
\bibitem [{\citenamefont {Martin}(2007)}]{Martin:2007equ}%
  \BibitemOpen
  \bibfield  {author} {\bibinfo {author} {\bibfnamefont {G.~D.~R.}\ \bibnamefont {Martin}},\ }\emph {\bibinfo {title} {{Classical and Quantum Radiation Reaction}}},\ \href@noop {} {Ph.D. thesis},\ \bibinfo  {school} {York U., England, Dept. Math.} (\bibinfo {year} {2007}),\ \Eprint {https://arxiv.org/abs/0805.0666} {arXiv:0805.0666 [gr-qc]} \BibitemShut {NoStop}%
\bibitem [{\citenamefont {Higuchi}\ and\ \citenamefont {Walker}(2009)}]{Higuchi:2009ms}%
  \BibitemOpen
  \bibfield  {author} {\bibinfo {author} {\bibfnamefont {A.}~\bibnamefont {Higuchi}}\ and\ \bibinfo {author} {\bibfnamefont {P.~J.}\ \bibnamefont {Walker}},\ }\bibfield  {title} {\bibinfo {title} {{Quantum corrections to the Larmor radiation formula in scalar electrodynamics}},\ }\href {https://doi.org/10.1103/PhysRevD.80.105019} {\bibfield  {journal} {\bibinfo  {journal} {Phys. Rev. D}\ }\textbf {\bibinfo {volume} {80}},\ \bibinfo {pages} {105019} (\bibinfo {year} {2009})},\ \Eprint {https://arxiv.org/abs/0908.2723} {arXiv:0908.2723 [hep-th]} \BibitemShut {NoStop}%
\bibitem [{\citenamefont {Tsai}(1986)}]{PhysRevD.34.1326}%
  \BibitemOpen
  \bibfield  {author} {\bibinfo {author} {\bibfnamefont {Y.~S.}\ \bibnamefont {Tsai}},\ }\bibfield  {title} {\bibinfo {title} {Axion bremsstrahlung by an electron beam},\ }\href {https://doi.org/10.1103/PhysRevD.34.1326} {\bibfield  {journal} {\bibinfo  {journal} {Phys. Rev. D}\ }\textbf {\bibinfo {volume} {34}},\ \bibinfo {pages} {1326} (\bibinfo {year} {1986})}\BibitemShut {NoStop}%
\bibitem [{\citenamefont {Aloni}\ \emph {et~al.}(2019)\citenamefont {Aloni}, \citenamefont {Fanelli}, \citenamefont {Soreq},\ and\ \citenamefont {Williams}}]{PhysRevLett.123.071801}%
  \BibitemOpen
  \bibfield  {author} {\bibinfo {author} {\bibfnamefont {D.}~\bibnamefont {Aloni}}, \bibinfo {author} {\bibfnamefont {C.}~\bibnamefont {Fanelli}}, \bibinfo {author} {\bibfnamefont {Y.}~\bibnamefont {Soreq}},\ and\ \bibinfo {author} {\bibfnamefont {M.}~\bibnamefont {Williams}},\ }\bibfield  {title} {\bibinfo {title} {Photoproduction of axionlike particles},\ }\href {https://doi.org/10.1103/PhysRevLett.123.071801} {\bibfield  {journal} {\bibinfo  {journal} {Phys. Rev. Lett.}\ }\textbf {\bibinfo {volume} {123}},\ \bibinfo {pages} {071801} (\bibinfo {year} {2019})}\BibitemShut {NoStop}%
\bibitem [{\citenamefont {Chen}\ and\ \citenamefont {Tajima}(1999)}]{PhysRevLett.83.256}%
  \BibitemOpen
  \bibfield  {author} {\bibinfo {author} {\bibfnamefont {P.}~\bibnamefont {Chen}}\ and\ \bibinfo {author} {\bibfnamefont {T.}~\bibnamefont {Tajima}},\ }\bibfield  {title} {\bibinfo {title} {Testing unruh radiation with ultraintense lasers},\ }\href {https://doi.org/10.1103/PhysRevLett.83.256} {\bibfield  {journal} {\bibinfo  {journal} {Phys. Rev. Lett.}\ }\textbf {\bibinfo {volume} {83}},\ \bibinfo {pages} {256} (\bibinfo {year} {1999})}\BibitemShut {NoStop}%
\bibitem [{\citenamefont {Avignone}\ \emph {et~al.}(1988)\citenamefont {Avignone}, \citenamefont {Baktash}, \citenamefont {Barker}, \citenamefont {Calaprice}, \citenamefont {Dunford}, \citenamefont {Haxton}, \citenamefont {Kahana}, \citenamefont {Kouzes}, \citenamefont {Miley},\ and\ \citenamefont {Moltz}}]{Avignone:1988bv}%
  \BibitemOpen
  \bibfield  {author} {\bibinfo {author} {\bibfnamefont {F.~T.}\ \bibnamefont {Avignone}}, \bibinfo {author} {\bibfnamefont {C.}~\bibnamefont {Baktash}}, \bibinfo {author} {\bibfnamefont {W.~C.}\ \bibnamefont {Barker}}, \bibinfo {author} {\bibfnamefont {F.~P.}\ \bibnamefont {Calaprice}}, \bibinfo {author} {\bibfnamefont {R.~W.}\ \bibnamefont {Dunford}}, \bibinfo {author} {\bibfnamefont {W.~C.}\ \bibnamefont {Haxton}}, \bibinfo {author} {\bibfnamefont {D.}~\bibnamefont {Kahana}}, \bibinfo {author} {\bibfnamefont {R.~T.}\ \bibnamefont {Kouzes}}, \bibinfo {author} {\bibfnamefont {H.~S.}\ \bibnamefont {Miley}},\ and\ \bibinfo {author} {\bibfnamefont {D.~M.}\ \bibnamefont {Moltz}},\ }\bibfield  {title} {\bibinfo {title} {{Search for Axions From the 1115-kev Transition of $^{65}$Cu}},\ }\href {https://doi.org/10.1103/PhysRevD.37.618} {\bibfield  {journal} {\bibinfo  {journal} {Phys. Rev. D}\ }\textbf {\bibinfo {volume} {37}},\ \bibinfo {pages} {618} (\bibinfo {year} {1988})}\BibitemShut {NoStop}%
\bibitem [{\citenamefont {Van~Tilburg}(2021)}]{PhysRevD.104.023019}%
  \BibitemOpen
  \bibfield  {author} {\bibinfo {author} {\bibfnamefont {K.}~\bibnamefont {Van~Tilburg}},\ }\bibfield  {title} {\bibinfo {title} {Stellar basins of gravitationally bound particles},\ }\href {https://doi.org/10.1103/PhysRevD.104.023019} {\bibfield  {journal} {\bibinfo  {journal} {Phys. Rev. D}\ }\textbf {\bibinfo {volume} {104}},\ \bibinfo {pages} {023019} (\bibinfo {year} {2021})}\BibitemShut {NoStop}%
\bibitem [{\citenamefont {Aprile}\ \emph {et~al.}(2022)\citenamefont {Aprile} \emph {et~al.}}]{PhysRevLett.129.161805}%
  \BibitemOpen
  \bibfield  {author} {\bibinfo {author} {\bibfnamefont {E.}~\bibnamefont {Aprile}} \emph {et~al.} (\bibinfo {collaboration} {XENON Collaboration}),\ }\bibfield  {title} {\bibinfo {title} {Search for new physics in electronic recoil data from xenonnt},\ }\href {https://doi.org/10.1103/PhysRevLett.129.161805} {\bibfield  {journal} {\bibinfo  {journal} {Phys. Rev. Lett.}\ }\textbf {\bibinfo {volume} {129}},\ \bibinfo {pages} {161805} (\bibinfo {year} {2022})}\BibitemShut {NoStop}%
\bibitem [{\citenamefont {Yan}\ \emph {et~al.}(2019)\citenamefont {Yan}, \citenamefont {Sun}, \citenamefont {Peng}, \citenamefont {Guo}, \citenamefont {Liu}, \citenamefont {Peng},\ and\ \citenamefont {Zheng}}]{Constraining_exotic_spin}%
  \BibitemOpen
  \bibfield  {author} {\bibinfo {author} {\bibfnamefont {H.}~\bibnamefont {Yan}}, \bibinfo {author} {\bibfnamefont {G.~A.}\ \bibnamefont {Sun}}, \bibinfo {author} {\bibfnamefont {S.~M.}\ \bibnamefont {Peng}}, \bibinfo {author} {\bibfnamefont {H.}~\bibnamefont {Guo}}, \bibinfo {author} {\bibfnamefont {B.~Q.}\ \bibnamefont {Liu}}, \bibinfo {author} {\bibfnamefont {M.}~\bibnamefont {Peng}},\ and\ \bibinfo {author} {\bibfnamefont {H.}~\bibnamefont {Zheng}},\ }\bibfield  {title} {\bibinfo {title} {Constraining exotic spin dependent interactions of muons and electrons},\ }\href {https://doi.org/10.1140/epjc/s10052-019-7442-8} {\bibfield  {journal} {\bibinfo  {journal} {The European Physical Journal C}\ }\textbf {\bibinfo {volume} {79}},\ \bibinfo {pages} {971} (\bibinfo {year} {2019})}\BibitemShut {NoStop}%
\bibitem [{\citenamefont {Zhitnitsky}(1980)}]{Zhitnitsky:1980tq}%
  \BibitemOpen
  \bibfield  {author} {\bibinfo {author} {\bibfnamefont {A.~R.}\ \bibnamefont {Zhitnitsky}},\ }\bibfield  {title} {\bibinfo {title} {{On Possible Suppression of the Axion Hadron Interactions. (In Russian)}},\ }\href@noop {} {\bibfield  {journal} {\bibinfo  {journal} {Sov. J. Nucl. Phys.}\ }\textbf {\bibinfo {volume} {31}},\ \bibinfo {pages} {260} (\bibinfo {year} {1980})}\BibitemShut {NoStop}%
\bibitem [{\citenamefont {Dine}\ \emph {et~al.}(1981)\citenamefont {Dine}, \citenamefont {Fischler},\ and\ \citenamefont {Srednicki}}]{Dine:1981rt}%
  \BibitemOpen
  \bibfield  {author} {\bibinfo {author} {\bibfnamefont {M.}~\bibnamefont {Dine}}, \bibinfo {author} {\bibfnamefont {W.}~\bibnamefont {Fischler}},\ and\ \bibinfo {author} {\bibfnamefont {M.}~\bibnamefont {Srednicki}},\ }\bibfield  {title} {\bibinfo {title} {{A Simple Solution to the Strong CP Problem with a Harmless Axion}},\ }\href {https://doi.org/10.1016/0370-2693(81)90590-6} {\bibfield  {journal} {\bibinfo  {journal} {Phys. Lett. B}\ }\textbf {\bibinfo {volume} {104}},\ \bibinfo {pages} {199} (\bibinfo {year} {1981})}\BibitemShut {NoStop}%
\bibitem [{\citenamefont {Raffelt}(1986)}]{PhysRevD.33.897}%
  \BibitemOpen
  \bibfield  {author} {\bibinfo {author} {\bibfnamefont {G.~G.}\ \bibnamefont {Raffelt}},\ }\bibfield  {title} {\bibinfo {title} {Astrophysical axion bounds diminished by screening effects},\ }\href {https://doi.org/10.1103/PhysRevD.33.897} {\bibfield  {journal} {\bibinfo  {journal} {Phys. Rev. D}\ }\textbf {\bibinfo {volume} {33}},\ \bibinfo {pages} {897} (\bibinfo {year} {1986})}\BibitemShut {NoStop}%
\bibitem [{\citenamefont {Kim}(1979)}]{PhysRevLett.43.103}%
  \BibitemOpen
  \bibfield  {author} {\bibinfo {author} {\bibfnamefont {J.~E.}\ \bibnamefont {Kim}},\ }\bibfield  {title} {\bibinfo {title} {Weak-interaction singlet and strong $\mathrm{CP}$ invariance},\ }\href {https://doi.org/10.1103/PhysRevLett.43.103} {\bibfield  {journal} {\bibinfo  {journal} {Phys. Rev. Lett.}\ }\textbf {\bibinfo {volume} {43}},\ \bibinfo {pages} {103} (\bibinfo {year} {1979})}\BibitemShut {NoStop}%
\bibitem [{\citenamefont {Shifman}\ \emph {et~al.}(1980)\citenamefont {Shifman}, \citenamefont {Vainshtein},\ and\ \citenamefont {Zakharov}}]{Shifman:1979if}%
  \BibitemOpen
  \bibfield  {author} {\bibinfo {author} {\bibfnamefont {M.~A.}\ \bibnamefont {Shifman}}, \bibinfo {author} {\bibfnamefont {A.~I.}\ \bibnamefont {Vainshtein}},\ and\ \bibinfo {author} {\bibfnamefont {V.~I.}\ \bibnamefont {Zakharov}},\ }\bibfield  {title} {\bibinfo {title} {{Can Confinement Ensure Natural CP Invariance of Strong Interactions?}},\ }\href {https://doi.org/10.1016/0550-3213(80)90209-6} {\bibfield  {journal} {\bibinfo  {journal} {Nucl. Phys. B}\ }\textbf {\bibinfo {volume} {166}},\ \bibinfo {pages} {493} (\bibinfo {year} {1980})}\BibitemShut {NoStop}%
\bibitem [{\citenamefont {Di~Piazza}(2021)}]{PhysRevD.103.076011}%
  \BibitemOpen
  \bibfield  {author} {\bibinfo {author} {\bibfnamefont {A.}~\bibnamefont {Di~Piazza}},\ }\bibfield  {title} {\bibinfo {title} {Wkb electron wave functions in a tightly focused laser beam},\ }\href {https://doi.org/10.1103/PhysRevD.103.076011} {\bibfield  {journal} {\bibinfo  {journal} {Phys. Rev. D}\ }\textbf {\bibinfo {volume} {103}},\ \bibinfo {pages} {076011} (\bibinfo {year} {2021})}\BibitemShut {NoStop}%
\bibitem [{cod(2025)}]{code}%
  \BibitemOpen
  \href {https://doi.org/https://doi.org/10.5281/zenodo.17406226} {}\bibinfo {howpublished} {\url{https://doi.org/10.5281/zenodo.17406226}} (\bibinfo {year} {2025})\BibitemShut {NoStop}%
\bibitem [{\citenamefont {Thomas}(1926{\natexlab{b}})}]{thomas1926motion}%
  \BibitemOpen
  \bibfield  {author} {\bibinfo {author} {\bibfnamefont {L.~H.}\ \bibnamefont {Thomas}},\ }\bibfield  {title} {\bibinfo {title} {The motion of the spinning electron},\ }\href@noop {} {\bibfield  {journal} {\bibinfo  {journal} {Nature}\ }\textbf {\bibinfo {volume} {117}},\ \bibinfo {pages} {514} (\bibinfo {year} {1926}{\natexlab{b}})}\BibitemShut {NoStop}%
\bibitem [{Note1()}]{Note1}%
  \BibitemOpen
  \bibinfo {note} {If we do not make the approximation $m_a \approx 0$, where $m_a$ is the axion mass, then $k_\perp ^2$ here is replaced by $k_\perp ^2 + m_a^2/\hbar ^2$.}\BibitemShut {Stop}%
\bibitem [{\citenamefont {Gradshteyn}\ and\ \citenamefont {Ryzhik}(2014)}]{gradshteyn2014table}%
  \BibitemOpen
  \bibfield  {author} {\bibinfo {author} {\bibfnamefont {I.~S.}\ \bibnamefont {Gradshteyn}}\ and\ \bibinfo {author} {\bibfnamefont {I.~M.}\ \bibnamefont {Ryzhik}},\ }\href@noop {} {\emph {\bibinfo {title} {Table of Integrals, Series, and Products - eighth edition}}}\ (\bibinfo  {publisher} {Academic Press},\ \bibinfo {year} {2014})\BibitemShut {NoStop}%
\end{thebibliography}

\providecommand{\noopsort}[1]{}\providecommand{\singleletter}[1]{#1}%

\end{document}